\documentclass[12pt]{article}
\usepackage{epsfig,psfig,graphicx,rotating}   
\begin{document}
\title{\bf
STATISTICAL MECHANICS OF THE SELF-GRAVITATING GAS WITH TWO OR MORE
KINDS OF PARTICLES}
\author{\bf  H. J. de Vega, J. A. Siebert \\ \\
Laboratoire de Physique Th\'eorique et Hautes Energies, \\
Universit\'e Paris VI, Tour 16, 1er \'etage, \\ 4, Place Jussieu
75252 Paris cedex 05, France. \\
Laboratoire Associ\'e au CNRS UMR 7589.}
\date{\today} 
\maketitle
\begin{abstract}
We study the statistical mechanics of the self-gravitating gas at thermal 
equilibrium with two kinds of particles. We start from the partition
function in the canonical ensemble which we express as a functional
integral over the densities of the two kinds of particles for a large
number of particles. The system is shown to possess an infinite volume limit 
when $(N_1,N_2,V) \to \infty$, keeping $ N_1 / V^{\frac{1}{3}} $
and $ N_2 / V^{\frac{1}{3}} $ fixed. The  saddle point
approximation becomes  here exact for $(N_1,N_2,V) \to \infty$. It
provides a nonlinear differential equation on the  densities of 
each  kind of particles. For the spherically symmetric case, we
compute the densities as functions of two dimensionless physical
parameters:  $\eta_1=\frac{G m_1^2 N_1}{V^{\frac{1}{3}}  \, T}$ and
$\eta_2=\frac{G m_2^2 N_2}{V^{\frac{1}{3}}  \, T}$
(where $G$ is Newton's constant, $m_1$ and $m_2$ the masses of
the two kinds of particles and $T$  the temperature). According to the
values of  $\eta_1$ and $\eta_2$  the system can be either in a
gaseous phase or in a highly condensed phase. The gaseous phase is
stable for  $\eta_1$ and $\eta_2$ between the origin and their
collapse values. We have thus generalized the well-known isothermal
sphere for two kinds of particles.
The gas  is inhomogeneous and the mass $M(R)$ inside a sphere of radius $R$ 
scales with $R$ as $M(R) \propto R^d $ suggesting a fractal
structure. The value of $d$ depends in general on  $\eta_1$ and
$\eta_2$ except on the critical line for the canonical ensemble in the
$(\eta_1,\eta_2)$ plane where it takes the universal value $ d \simeq
1.6 $ for all values of $ N_1/N_2 $. The equation of state is
computed. It is found to be {\bf locally} a perfect gas equation of
state. The thermodynamic functions (free energy, energy, entropy) are expressed
and plotted as functions of $\eta_1$ and $\eta_2$.  They 
exhibit a square root Riemann sheet with the branch points on the critical
canonical line. The behaviour of the energy and the specific
heat at the critical line is computed.
This treatment is further generalized to the self-gravitating gas with
$n$-types of particles.
\end{abstract} 
\newpage

\tableofcontents

\section{Introduction}   

The self-gravitating gases have remarkable physical properties due to
the long range nature of the gravitational force. They are not
homogeneous even at thermal equilibrium. This  fundamental
inhomogeneity character suggested that fractal structures can arise in
a self-interacting gravitational gas\cite{pre,pcm,1sg}. 

Self-gravitating gases are used to describe cold clouds in the
interstellar medium as well as the large scale structure of
galaxies. In both cases, self-gravitating gases provide scaling laws
for the mass distribution with Haussdorf dimensions compatible with
the observations\cite{pre,pcm,1sg}.

All particles have the same mass in the self-gravitating gases in
thermal equilibrium considered till now both in the hydrostatic
approach \cite{hidro} and in the statistical mechanics approach\cite{1sg}.
We study in the present paper the statistical mechanics of a
non-relativistic gas with two kinds of particles with
masses $m_1$ and $m_2$. Such a system, besides its own physical
interest, has obvious astrophysical motivations since cold clouds in
the galaxy are formed typically by several kinds of particles. For
example, hydrogen and helium. 

\bigskip

Let us begin by briefly recalling some results about the statistical
mechanics of  the self-gravitating gas with one kind of particles \cite{1sg}.
It is a gas of $N$ non-relativistic particles
of mass $m$ interacting through Newtonian gravity. The gas is in a volume
$V$ and in a thermal bath at temperature $T$. In the usual 
thermodynamic limit ($N, \, V \to \infty$ and $ N /V $ is fixed)
the gaseous phase is not stable and the system collapses in a very
dense phase. In the dilute limit 
$N, \, V \to \infty$ and $ N/V^{\frac13}$ fixed the system
can exist in the gaseous phase. This is a dilute limit since the
average density  $\frac{N}{V}$ goes as $ V^{-\frac{2}{3}} \to 0 $.
The relevant physical parameter of the
system is $\eta=\frac{G m^2 N}{L T}$ with Newton's constant $G$ and
the  length $ L \equiv V^{\frac{1}{3}}$. $\eta$ is the ratio 
of the characteristic gravitational energy $\frac{G m^2 N}{L}$ and the
kinetic energy $T$ of the gas. For $\eta=0$ the ideal gas is recovered.
From $\eta=0$ till a critical value $\eta_0 = 2.43450\ldots$
the gaseous phase is stable in the canonical ensemble. When $\eta$ reaches the 
value $\eta_0$ the gas collapses in a very dense phase. The velocity of
sound becomes imaginary at this point triggering unstabilities that
lead to the collapse of the gas\cite{1sg}. The saddle point
approximation applies between $\eta=0$ and the point $\eta^C = 2.517551\ldots$ 
(associated to the Jeans instability)
where the determinant of small fluctuations is positive. At $\eta^C$ 
the determinant of small fluctuations vanishes and the saddle point
approximation breaks down\cite{1sg}. Beyond $\eta^C$ the gaseous phase
is stable and the mean field approximation holds in the microcanonical
ensemble. Solving the saddle point  equation in the spherically 
symmetric case allows to obtain the particle density and the
thermodynamic functions as functions of the physical parameter
$\eta$. As shown in ref.\cite{1sg} the mean field approach is
equivalent to the hydrostatic description\cite{hidro} provided the
ideal gas equation of state is postulated in the latter approach.
The mass distribution turns to exhibit a scaling
behaviour as a function of $R$\cite{1sg}.

\bigskip

We consider in this paper the  self-gravitating gas with two kinds of
particles in the canonical ensemble. We recast the 
partition function as a functional integral over the densities of
particles of each kind when the number of particles is large.
The statistical weight for each configuration
of densities turns to be the exponential of an `effective action'
which is proportional to the number of particles. Therefore,
we can use the saddle point approximation in the thermodynamic limit 
to evaluate the partition function. The `effective
action' turns to be the free energy as a functional of the particle densities.

When the saddle point provides a minimum of the free energy,
the density solution of the saddle point equation is the most probable.
It is  certainly exact for an infinite number of particles, since
the minimized free energy  exponentially dominates the
partition function. That is, the mean field theory defined by the saddle point
becomes exact for an infinite number of particles. The mean field
approximation for the canonical ensemble ceases to be valid on a
critical line in the $ (\eta_1, \eta_2) $ plane (see below). Beyond
this critical line the saddle point is not a minimum of the free
energy and it fails to reproduce the physics of the system.

\bigskip

More precisely, we consider $N_1$ particles of mass 
$m_1$ and $N_2$ particles of mass $m_2$ interacting through Newtonian 
gravity in a volume $V$ and in a thermal equilibrium at temperature
$T$. By analogy with the self-gravitating gas with one kind of
particles, we consider the dilute thermodynamic limit:
\begin{equation} \label{limter}
N_1, \; N_2, \; V \to \infty \quad \mbox{keeping}\quad  
 \frac{N_1}{V^{\frac{1}{3}}}\quad \mbox{and}\quad \frac{N_2}{V^{\frac{1}{3}}}
\quad   \mbox{fixed} \; 
\end {equation}
The two relevant physical parameter are here 
\begin{equation} \label{etai}
\eta_1=\frac{G m_1^2 N_1}{L \;  T} \quad , \quad
\eta_2=\frac{G m_2^2 N_2}{L \; T} 
\end {equation}
where $ L \equiv V^{\frac{1}{3}} $ for a cubic geometry. 
Notice that eq.(\ref{limter}) implies that the ratio
$N_1/N_2$ stays fixed for $ N_1, \; N_2, \; V \to \infty .$

The self-gravitating gas with two kinds of particle behaves as a
perfect gas in the extremely diluted limit  $\eta_1 \to 0 $ and
$\eta_2 \to 0 $. When $ \eta_1 $ and/or $ \eta_2 $ grow, the gas
becomes denser till it collapses into a very dense phase when
$\eta_1$ and $\eta_2$ reach the collapse line for the canonical
ensemble in the $ (\eta_1,\eta_2) $ plane. By analogy with the gas
with one kind of particles \cite{1sg} we expect the collapse line to
be very close and below the critical line in the $ (\eta_1,\eta_2) $ plane.
The gaseous phase keeps stable in the microcanonical ensemble beyond
this critical line. 

We find here that the saddle point equations are two coupled
non-linear differential equations for
the densities of the two kinds of particles: $\rho_1({\bf x}),\rho_2({\bf x})$.
We succeed to reduce these equations to a single 
non-linear differential equation. We solve it in the spherically 
symmetric case. We express the densities as functions of the physical
parameters $\eta_1$ and $\eta_2$. We thus find the isothermal sphere
with two types of particles.

We compute the mass inside a sphere
of radius $ R $ centered at the origin and show that it scales with a
Haussdorf dimension $d$. $d$ decreases with $\eta_1^R$ and $\eta_2^R$
from the value $d=3$ for the ideal homogeneous gas
till $ d\approx 1.6 $ in the canonical critical line. 
The  Haussdorf dimension  keeps decreasing beyond the canonical
critical line in the stable phase of the microcanonical ensemble. 
$d$ at the canonical critical line turns to be {\bf independent} of the ratio
$ \eta_1^R / \eta_2^R $ and  coinciding within the numerical
precision with $ d \approx 1.6 $ for  the
canonical critical point of the gas with one kind of particles \cite{1sg}.
This indicates that the  Haussdorf dimension at  the
canonical critical line is an universal value, independent of the gas
composition. 

The dependence of the critical values of the parameters $\eta_1$ and $\eta_2$
with the number of particles ratio is computed.
The thermodynamic functions (free energy, energy, entropy,
local pressure and pressure contrast) are expressed as functions  of
the physical parameter $\eta_1$ and $\eta_2$. The pressure contrast
[ratio between the pressure at the origin and the pressure at
the boundary] turns to be {\bf lower} for this mixture of particles than for
the gas with one kind of particles.

We compute the  pressure at a point $ \bf r $ of the gas and show
that it {\bf locally} obeys the equation of state of an ideal gas,
\begin{equation} \label{ecestl}
P({\bf r})=\frac{N_1 T}{V} \; \rho_1({\bf r}) +
\frac{N_2 T}{V} \; \rho_2({\bf r}) \;.
\end{equation} 
Since the gas is inhomogeneous, the pressure acting on any
non-infinitesimal volume of the gas does not obey the ideal
equation of state. (This was already the case for a gas with all particles of
equal mass\cite{1sg}). We plot in figs. \ref{pressure3d}-\ref{pressuresec}
the pressure of the gas at the surface. 

The mean field equations have a straightforward hydrostatic
interpretation. We  show that the mean field equations derived from
the partition function are equivalent to the hydrostatic equilibrium
equations {\bf provided} that the ideal equation of state is
postulated in the latter approach. We stress that we give 
here a {\bf microscopic derivation} of 
the equation of state (\ref{ecestl}) from the partition function.

We then consider a self-gravitating gas formed by $n$ kinds of
particles with different masses. The mean field equations are derived
and shown to reduce to a single non-linear differential equation.

This paper is organized as follows,  In section II
we present the statistical mechanics of the self-gravitating gas with
two kinds of particles in the canonical ensemble, in sec. III we
present the main thermodynamic magnitudes, the equation of state and
the scaling behaviour of the particle distribution for spherical symmetry.
In sec. IV we present the generalization for $n$ kinds of
particles. The Appendixes contain relevant mathematical developments
and the hydrostatic approach to a self-gravitating gas with two kinds
of particles. 

\section{Statistical mechanics and mean field theory 
for the self-gravitating gas }

We present here the partition function for the self-gravitating gas
with $N_1$ particles of mass $m_1$ and $N_2$ particles of mass $m_2$
inside a finite volume $V$ and derive the mean field approach to it. 

\subsection{The canonical ensemble}

We study the statistical mechanics of the self-gravitating gas 
with two kinds of particles
in the canonical ensemble. The Hamiltonian of the system is 
\begin {eqnarray}
\label{Ham}
&&H= \sum_{i=1}^{N_1} \frac{{\bf p^2_{1,i}}}{2 m_1} \; + \;
\sum_{i=1}^{N_2} \frac{{\bf p^2_{2,i}}}{2 m_2} \nonumber \\ 
&&-\sum_{1 \leq i < j \leq N_1} \frac {G \;  m_1^2} {|{\bf q_{1,i}}-{\bf
q_{1,j}}|} 
-\sum_{1 \leq i < j \leq N_2} \frac {G  \; m_2^2} {|{\bf q_{2,i}}-{\bf
q_{2,j}}|} 
-\sum_{1 \leq  i \leq N_1,1 \leq  j \leq N_2    } 
\frac {G  \; m_1 m_2} {|{\bf q_{1,i}}-{\bf
q_{2,j}}|}  \; \;.\nonumber  
\end{eqnarray}   
\noindent Here, ${\bf p_{1,i}}$ and ${\bf q_{1,i}}$ are the momenta and the
coordinates of 
the particles of mass $m_1$. ${\bf p_{2,i}}$ and ${\bf q_{2,i}}$ 
are the momenta and the coordinates of the particles of mass $m_2$.
Therefore, the classical partition function of the gas is 
$$
Z(T,N_1,N_2,V)=\frac{1}{N_1!N_2!} \int\ldots\int \prod_{l=1}^{l=N_1}
\frac{{\rm d}^3 {\bf p_{1,l}} \; {\rm d}^3 {\bf q_{1,l}}}{(2 \pi)^3}
\prod_{l=1}^{l=N_2}
\frac{{\rm d}^3 {\bf p_{2,l}} \; {\rm d}^3 {\bf q_{2,l}}}{(2 \pi)^3} \;
e^{- \frac{H}{T} } \; \;.
$$
\noindent
It is convenient to introduce the dimensionless coordinates variables
${\bf r_{1,l}}=L \; {\bf q_{1,l}}$ and ${\bf r_{2,l}}=L \; {\bf q_{2,l}}$.
The momenta integrals are computed straightforwardly. Hence the
partition function becomes the product of the partition function of 
perfect gases with masses $m_1$ and $m_2$ times the coordinate
integral $Z_{int}$. 
$$
Z=\frac {V^{N_1}}{N_1!} \left(\frac {m_1 T}{2 \pi}\right)^{3 N_1/2}
 \;  \frac {V^{N_2}}{N_2!} \left(\frac {m_2 T}{2 \pi}\right)^{3 N_2/2}
 \; Z_{int} \; ,
$$
where
$$
Z_{int}=\int\ldots\int \prod_{l=1}^{l=N_1}{\rm d}^3{\bf r_{1,l}}
\prod_{k=1}^{k=N_2}{\rm d}^3{\bf r_{2,k}} \;
\exp{\left( \eta_1  \, u_{11} +\eta_2 \, u_{22} +\sqrt{\eta_1 \eta_2}\,
u_{12} \right)} \; ,
$$
and
\begin{eqnarray} 
u_{11}& \equiv &\frac {1}{N_1} \sum_{1 \leq i < j \leq N_1} 
\frac {1}{|{\bf r_{1,i}}-{\bf r_{1,j}}|} \quad , \quad
u_{22}\equiv\frac {1}{N_2} \sum_{1 \leq i < j \leq N_2} 
\frac {1}{|{\bf r_{2,i}}-{\bf r_{2,j}}|} \quad , \nonumber\\
u_{12}&\equiv&\frac {1}{\sqrt{N_1 N_2}} \sum_{1 \leq i \leq N_1,1 \leq
j \leq N_2 } \frac {1}{|{\bf r_{1,i}}-{\bf r_{2,j}}|} \; . \nonumber
\end{eqnarray}

\subsection{Mean field theory}

We approximate the function $Z_{int}$  for a large number of particles
($N_1\gg1$ and $N_2\gg1$) generalizing  the approach of
ref.\cite{1sg,Lipatov} for two kinds of particles. The function $Z_{int}$
is expressed as a functional integral over all configurations with particle 
densities $\rho_1({\bf x})$ and $\rho_2({\bf x})$
[$\rho_1({\bf x})$ stands for the density of the particles of mass $m_1$
and $\rho_2({\bf x})$ stands for the density of the particles of mass $m_2$].
\begin {equation}
\label{intfonct}
Z_{int}=\int D\rho_1(.) \; D\rho_2(.) \; {\rm d}b_1 \; {\rm d}b_2 \;
 e^{-\frac{F-F_0}{T}}
\end{equation}
\noindent
with,
\begin {eqnarray}
\label{action}
\frac{F-F_0}{T}&=&
-N_1 \; \frac{\eta_1}{2}\int\frac{ {\rm d}^3{\bf x} \, {\rm d}^3{\bf y}}
{|{\bf x}-{\bf y}|} \; \rho_1({\bf x})\; \rho_1({\bf y})
-N_2 \; \frac{\eta_2}{2}\int\frac{ {\rm d}^3{\bf x} \, {\rm d}^3{\bf y}}
{|{\bf x}-{\bf y}|} \; \rho_2({\bf x})\; \rho_2({\bf y})\nonumber\\
&&-\sqrt{N_1 N_2} \; \sqrt{\eta_1 \eta_2}
\int\frac{ {\rm d}^3{\bf x} \, {\rm d}^3{\bf y}}
{|{\bf x}-{\bf y}|} \; \rho_1({\bf x})\; \rho_2({\bf y})+ \nonumber\\
&&+N_1 \int {\rm d}^3 {\bf x} \; \rho_1({\bf x}) \ln{\rho_1({\bf x} )}
+N_2\int{\rm d}^3{\bf x} \;\rho_2({\bf x}) \ln{\rho_2({\bf x} )} +\nonumber\\
&& +i N_1 \;  b_1 [1-\int {\rm d}^3 {\bf x} \; \rho_1({\bf x})]
 +i N_2 \;  b_2 [1-\int {\rm d}^3 {\bf x} \; \rho_2({\bf x})] \; \; .
\end{eqnarray}
\noindent
$b_1$ and $b_2$ are Lagrange multiplier enforcing the normalization of the 
densities: 
\begin {equation} \label{normalisation}
\int {\rm d}^3 {\bf x}\;\rho_1({\bf x})=1 \quad , \quad 
\int {\rm d}^3 {\bf x}\;\rho_2({\bf x})=1   \; \; .
\end{equation}
\noindent $F$ stands for the free energy of the gas for
the pair of densities $(\rho_1,\rho_2)$, while 
$$ 
F_0=-N_1 T \ln{\left[\frac{e V}{N_1} \left(\frac{m_1 T}{2 \pi}\right)^
{\frac{3}{2}}\right]}
-N_2 T \ln{\left[\frac{e V}{N_2} \left(\frac{m_2 T}{2 \pi}\right)^
{\frac{3}{2}}\right]} 
$$ 
is the free energy of the perfect gas with masses  $m_1$ and $m_2$.
One recognizes  the gravitational energy in the first two lines 
of eq.(\ref {action}), while the third line contains the entropy.

\vspace{1 cm}

Since the free energy becomes large in the 
thermodynamic limit ($N_1,N_2\gg1$), the functional integral 
$Z_{int}$ (\ref{intfonct}) is dominated by the minima of $\frac{F-F_0}{T}$.
Extremizing
the free energy with respect to the pair of densities $(\rho_1,\rho_2)$
yields the saddle point equations
\begin {eqnarray} \label{sadeq}
\ln{\rho_1({\bf x})}&=&a_1+\eta_1 \int
\frac{{\rm d}^3{\bf y}}{|{\bf y}-{\bf x}|}\rho_1({\bf y})
+\mu \;  \eta_2 \int
\frac{{\rm d}^3{\bf y}}{|{\bf y}-{\bf x}|}\rho_2({\bf y}) \\
\ln{\rho_2({\bf x})}&=&a_2
+\frac{1}{\mu} \; \eta_1 \int
\frac{{\rm d}^3{\bf y}}{|{\bf y}-{\bf x}|}\rho_1({\bf y})
+\eta_2 \int
\frac{{\rm d}^3{\bf y}}{|{\bf y}-{\bf x}|}\rho_2({\bf y}) \;\;.\nonumber
\end{eqnarray}
\noindent where we used eq.(\ref{etai}). These equations define the
mean field approach. We set $a_1=-1+i b_1$ and  $a_2=-1+i b_2$ and we
denote by $\mu$ the ratio of masses of the two kinds of particles,
$$ 
\mu \equiv \frac{m_1}{m_2}\;.
$$ 
\noindent 
We set,
\begin {equation} \label{phi}
\rho_1({\bf x})=\exp[\Phi_1({\bf x})]\quad , \quad
\rho_2({\bf x})=\exp[\Phi_2({\bf x})] \; .
\end{equation}
\noindent Eqs.(\ref{sadeq}) give for the gravitational potential
\begin{equation}   \label{potentiel}
U({\bf x})=-\frac{T}{m_1}[\Phi_1({\bf x})-a_1]
=-\frac{T}{m_2}[\Phi_2({\bf x})-a_2] 
\end{equation} 
\noindent
created by the matter densities $\rho_1(.)$ and $\rho_2(.)$ at the
point $ {\bf x} $. Using eqs.(\ref{phi})-(\ref{potentiel}), we see that
these densities obey the Boltzmann laws
\begin {equation} \label{Boltzmann}
\rho_1({\bf x})=e^{a_1} \; e^{-\frac{m_1}{T}U({\bf x})} \quad , \quad
\rho_2({\bf x})=e^{a_2} \; e^{-\frac{m_2}{T}U({\bf x})} \; ,
\end{equation}
\noindent 
containing the energy of a particle 
in this  mean field gravitational potential, as it must be.
$e^{a_1}$ and $e^{a_2}$ play the role of normalization constants.

\noindent 
Applying the Laplace operator to the saddle point equations (\ref{sadeq})
we find the differential equations
\begin {eqnarray} \label{Poisson}
\Delta \Phi_1({\bf x})+4 \pi \; \eta_1 \; e^{\Phi_1({\bf x})}+
4 \pi \; \mu \; \eta_2  \; e^{\Phi_2({\bf x})}
&=&0 \nonumber \\
\Delta \Phi_2({\bf x})+4 \pi  \; \frac{\eta_1}{\mu} \; e^{\Phi_1({\bf x})}+
4 \pi \;  \eta_2 \;  e^{\Phi_2({\bf x})}
&=&0\;.
\end{eqnarray}
These equations are scale covariant. If $(\Phi_1,\Phi_2)$ is
a pair of solutions of eqs.(\ref {Poisson}), then
the pair  $(\Phi_{1 \lambda},\Phi_{2 \lambda})$ defined by
\begin{equation} \label{scaletrans}
\Phi_{1 \lambda}({\bf x})=\Phi_1(\lambda {\bf x})+\ln{\lambda^2} \quad , \quad
\Phi_{2 \lambda}({\bf x})=\Phi_2(\lambda {\bf x})+\ln{\lambda^2} 
\end{equation}
is also a solution of eq.(\ref {Poisson}). This property is due to
the scale behaviour of Newton's potential.
\noindent
Using eq.(\ref{potentiel}), we reduce eqs.(\ref{Poisson}) to a single equation
\begin {equation} \label{Poired}
\Delta \Phi_1({\bf x})+4 \pi \; \eta_1 \; e^{\Phi_1({\bf x})}+
4 \pi \mu \; \eta_2 \; e^{a_2-\frac{a_1}{\mu}} \;
e^{\frac{\Phi_1({\bf x})}{\mu}} =0 \; .
\end{equation}
Using eqs.(\ref{potentiel}) and (\ref{Boltzmann}), eq.(\ref{Poired})
becomes in dimensionless coordinates,
\begin {equation} \label{hidro}
\Delta U({\bf x})= \frac{4 \, \pi \, G}{L} \left[ m_1 \; N_1
\;\rho_1({\bf x}) + m_2 \; N_2 \;\rho_2({\bf x})\right] \; .
\end{equation}
We show in Appendix B that this equation is the condition of hydrostatic 
equilibrium  for a two-component fluid once  the ideal gas equation of
state is postulated. 

Therefore, the mean field approximation is equivalent to the
hydrostatic description in the gaseous phase provided the ideal gas
equation of state is assumed in the latter approach.

Notice that local equations of state other than the ideal gas are
often assumed in the context of self-gravitating
fluids\cite{hidro}. As stressed in ref.\cite{1sg}, one needs long
range forces other than gravitational in order to 
obtain a non-ideal local equation of state in thermal equilibrium. 

\section{Spherically symmetric case}

For spherically symmetric configurations the mean field equations
become ordinary non-linear differential equations. We express here the
various thermodynamic quantities in terms of the solution of a single
ordinary differential equation. Such equation
reduces to the well-known isothermal sphere\cite{hidro} if all 
particles have identical mass.

\subsection{Reduction of the equation of saddle point}

We consider here the spherically symmetric case where the mean field
equations (\ref {Poisson})  take the form
\begin {eqnarray} \label{Poissonrad}
\frac{{\rm d}^2 \Phi_1}{{\rm d}R^2}+\frac{2}{R} \frac {{\rm d} \Phi_1}{{\rm
d}R} +4 \pi \; \eta_1 \; e^{\Phi_1(R)}+ 4\pi \; \mu  \; \eta_2 \;
e^{\Phi_2(R)} =0 \\ \frac{{\rm d}^2 \Phi_2}{{\rm d}R^2}+\frac{2}{R}
\frac {{\rm d} \Phi_2}{{\rm d}R}+4 \pi \;  \frac{\eta_1}{\mu} \;
e^{\Phi_1(R)}+4\pi \;  \eta_2 \; e^{\Phi_2(R)} =0 \nonumber
\end{eqnarray}
\noindent 
where we work in a unit sphere. Therefore the radial variable runs
in the interval $0 \leq R \leq R_{max} $,
$ R_{max}=\left(\frac{3}{4 \pi}\right)^{\frac{1}{3}} $.
Using the scale covariance of the mean field eqs. (\ref {Poisson})
by the transformation (\ref {scaletrans}), we can set
\begin {eqnarray} \label{dilatation}
\Phi_1(R)&=&\chi_1\left(\lambda \; \frac{R}{R_{max}}\right)
+\ln\left(\frac{\lambda^2}{3 \eta_1^R}\right)\nonumber \\
\Phi_2(R)&=&\chi_2\left(\lambda \; \frac{R}{R_{max}}\right)
+\ln\left(\frac{\lambda^2}{3 \eta_2^R}\right)\;.
\end{eqnarray}
\noindent 
We use the new parameters $ \eta_1^R= \eta_1/R_{max} $ and
$\eta_2^R=\eta_2/R_{max} $ in the spherically symmetric case.
Hence the mean field eqs.(\ref {Poissonrad}) are transformed 
into  a reduced system of the form,
\begin {eqnarray}\label{Poissonchi}
\chi_1^{''}(\lambda)+\frac{2}{\lambda} \,
\chi_1^{'}(\lambda)+e^{\chi_1(\lambda)} +\mu \; e^{\chi_2(\lambda)}
&=& 0\nonumber  \\
\chi_2^{''}(\lambda)+\frac{2}{\lambda} \, \chi_2^{'}(\lambda)+
\frac{1}{\mu}\; e^{\chi_1(\lambda)} +e^{\chi_2(\lambda)}
&=&0  \; .
\end{eqnarray}
\noindent 
Let us find the boundary conditions of these  equations.
In order to have a regular solution at origin we impose
$$ 
\chi_1^{'}(0)=0 \quad , \quad
\chi_2^{'}(0)=0 \; .
$$ 
\noindent 
The system (\ref{Poissonchi}) is invariant under the transformation
$$ 
\lambda \to \lambda \; e^{\alpha} \quad , \quad \chi_i \to \chi_i-2
\alpha \quad i = 1, 2 \;.
$$ 
\noindent
Hence, we can choose 
\begin{equation} \label{limcond2}
\chi_1(0)=0\; .
\end{equation}
\noindent
without loosing generality.
\noindent
As we see below, the remaining boundary condition on $\chi_2(0)$
is not independent from $\eta_1$ and $\eta_2$.

The densities of the two kinds of particles  $\rho_1$ and $\rho_2$ are
to be normalized according to eq.(\ref {normalisation}). We obtain using
eqs.(\ref{phi}) and (\ref{dilatation}), 
\begin{equation}\label{etai-norm}
\eta_1^R=\frac{1}{\lambda}
\int_{0}^{\lambda} {\rm d}x \; x^2  \; e^{\chi_1(x)} \quad , \quad
\eta_2^R=\frac{1}{\lambda}
\int_{0}^{\lambda} {\rm d}x \;  x^2 \;  e^{\chi_2(x)} \; .
\end{equation}
\noindent
Using the reduced mean field equations (\ref {Poissonchi}), it is
straightforward to show  from eq.(\ref{etai-norm}) that
\begin {eqnarray} \label{gauss}
\eta_1^R+\mu \; \eta_2^R&=&-\lambda \chi_1^{'}(\lambda) \\
\frac{1}{\mu} \; \eta_1^R+\eta_2^R&=&-\lambda \chi_2^{'}(\lambda) \; .\nonumber
\end{eqnarray}
\noindent
Hence
\begin {equation}  \label{champ}
\chi_2^{'}(\lambda)=\frac{1}{\mu} \; \chi_1^{'}(\lambda) \;.
\end{equation}  
\noindent
Recalling that $U$ is the  gravitational potential (\ref {potentiel})
we see using eqs.(\ref{dilatation}), that
$$
\frac{T}{m_1} \; \chi_1^{'}(\lambda)=\frac{T}{m_2} \; \chi_2^{'}(\lambda)
$$
is the gravitational field at the boundary of the sphere
($ R=R_{max} $) consistently with eq.(\ref{champ}).
From eq.(\ref {champ})  we introduce a new parameter
\begin {equation} \label{constante}
C \equiv \chi_2(\lambda)-\frac{1}{\mu} \; \chi_1(\lambda)
\end{equation}
\noindent
independent of $\lambda$ and function   only of the physical
parameters $\eta_1^R$ and $\eta_2^R$. Notice the boundary condition
\begin {equation} \label{qui20}
\chi_2(0)=C \; .
\end{equation}
The reduced mean field equations
(\ref {Poissonchi}) become a single equation with the parameter $C$
as coefficient
\begin{equation} \label{eqchi1}
\chi_1^{''}(\lambda)+\frac{2}{\lambda}\,
\chi_1^{'}(\lambda)+e^{\chi_1(\lambda)} 
+\mu \;  e^{C} \; e^{\frac{1}{\mu} \, \chi_1(\lambda)} =0
\end{equation}
and  the boundary  conditions $ \chi_1(0)=0 , \; \chi_1^{'}(0)=0 $.

Using eqs.(\ref{phi}) and (\ref {dilatation}) we can express the
densities of the two 
kinds of particle in terms of the solution of eq.(\ref {eqchi1})
\noindent
\begin{equation} \label{dens}
\rho_1(R)=\frac{\lambda^2}{3 \eta_1^R} \;
e^{\chi_1\left(\lambda\frac{R}{R_{max}}  \right)} \quad , \quad
\rho_2(R)=\frac{\lambda^2 \; e^{C} }{3 \eta_2^R} \;
e^{\frac{1}{\mu}\chi_1\left(\lambda \frac{R}{R_{max}} \right)}\;\;.
\end{equation}
\noindent
Both $\chi_1(\lambda)$ and $\chi_2(\lambda)$ are functions  of $C$
[see (\ref {constante})-(\ref {eqchi1})].
Inserting $\chi_1(x)$ and $\chi_2(x)$ in eq.(\ref {etai-norm}),
we see that $\eta_1^R$ and $\eta_2^R$ become functions of
$\lambda$ and $C$. Then expressing $\lambda$ and $C$ as functions of
$\eta_1^R$ and $\eta_2^R$, both densities of particles
in eq.(\ref{dens}) become functions 
of the radial variable $R$ and of the physical parameters
$\eta_1^R$ and $\eta_2^R$. 

We find asymptotically from eq.(\ref{eqchi1}) 
$$
\chi_1(\lambda)\buildrel{\lambda \to \infty }\over= - 2\, \mu\; \log \lambda +
{\cal O}(1) 
$$
for $ \mu \geq 1 $. Notice that this asymptotic behaviour may apply
for non-physical values of $\lambda$ where the gas is actually collapsed.

We now compute the thermodynamic quantities as functions of the
physical parameters $\eta_1^R$ and $\eta_2^R$.

\subsection{Free energy}

\begin{figure}[htbp]
\rotatebox{-90}{\epsfig{file=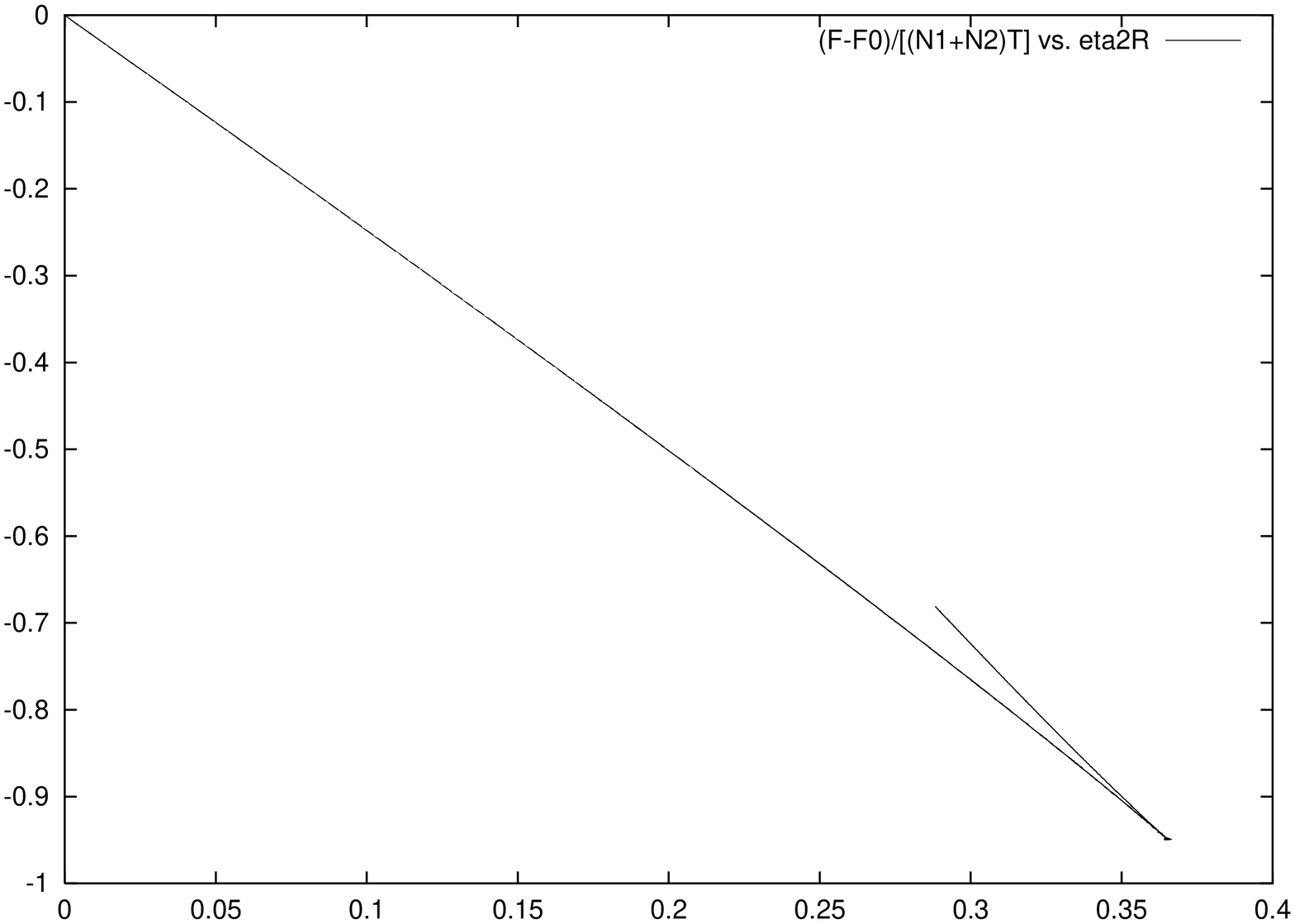,width=14cm,height=14cm}}
\caption{Free energy versus $\eta_2^R$ for 
$\frac{m_1}{m_2}=4, \; \frac{N_1}{N_2}=\frac{1}{3}$ and therefore  
$\frac{\eta_1^R}{\eta_2^R}=\frac{16}{3}$.} 
\label{freeen}
\end{figure}

\begin{figure}[htbp]
\rotatebox{-90}{\epsfig{file=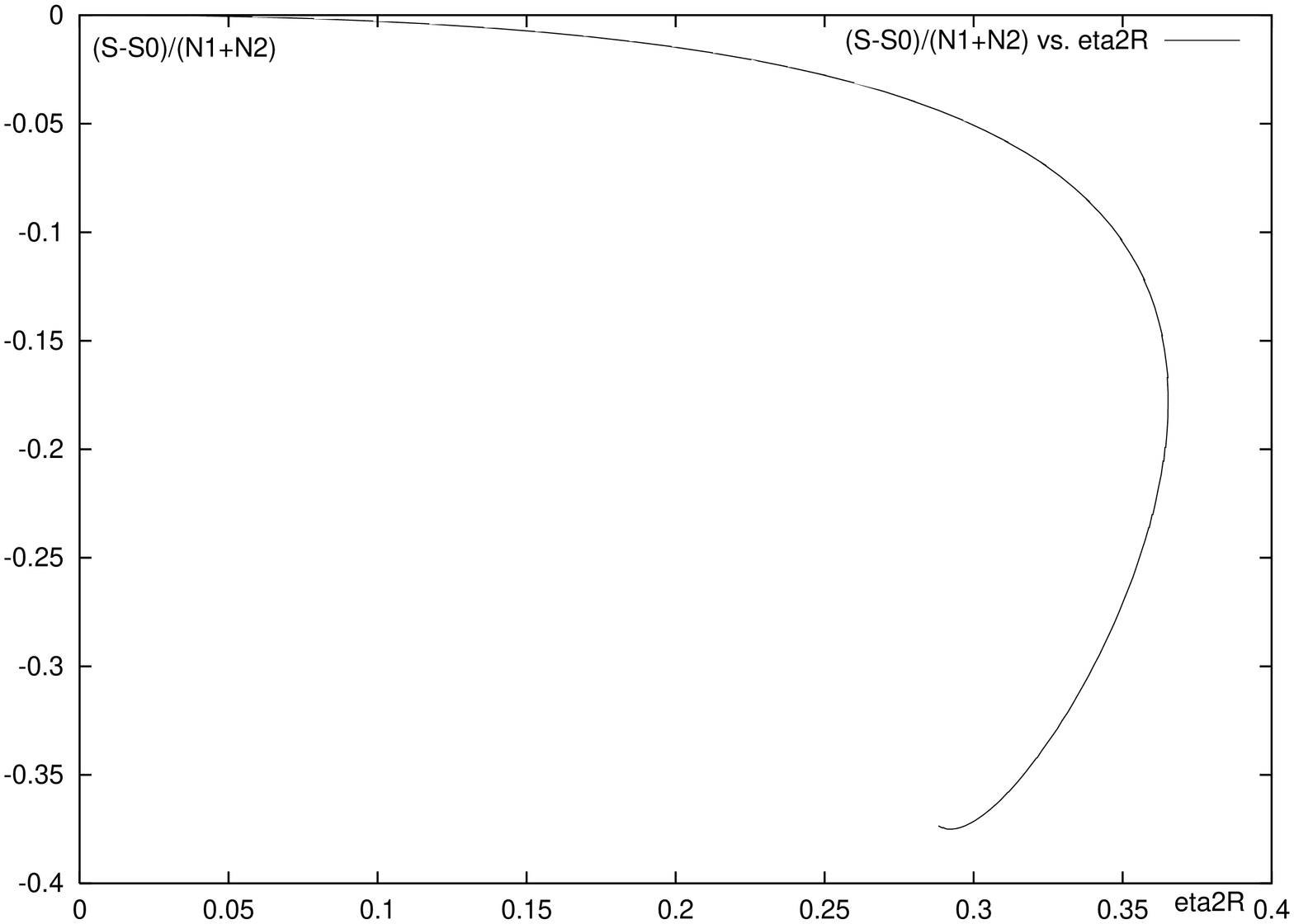,width=14cm,height=14cm}}
\caption{Entropy versus $\eta_2^R$ for 
$\frac{m_1}{m_2}=4, \; \frac{N_1}{N_2}=\frac{1}{3}$ and therefore  
$\frac{\eta_1^R}{\eta_2^R}=\frac{16}{3}$.} 
\label{entropy}
\end{figure}
\noindent
Let us start by computing the free energy.
Using eqs. (\ref {action})-(\ref{phi}) we find 
\begin {equation}
\label{free1}
\frac{F-F_0}{T}=\frac{N_1}{2}\left(a_1+\int {\rm d}^3{\bf x} \;
\Phi_1({\bf x}) \; e^{\Phi_1({\bf x})} \right)+
\frac{N_2}{2}\left(a_2+\int {\rm d}^3{\bf x} \; \Phi_2({\bf x}) \;
e^{\Phi_2({\bf x})} \right) \;.
\end{equation}
\noindent
We express now the Lagrange multipliers $a_1$ and $a_2$  as functions of the 
physical parameters $\eta_1^R$ and $\eta_2^R$. In the spherically
symmetric case the integration over the angles in eqs.(\ref {sadeq}) yields
\begin {eqnarray} \label{intangle}
\Phi_1(R)&=&a_1+4 \pi\eta_1\left(\frac{1}{R}\int_{0}^{R} {\rm d}R^{'}
\; R^{'2} \; e^{\Phi_1(R^{'})} +\int_{R}^{R_{max}}{\rm d}R^{'} \;
R^{'} \; e^{\Phi_1(R^{'})} \right)\nonumber \\
&&+ 4 \pi \; \mu \, \eta_2\left(\frac{1}{R}\int_{0}^{R} {\rm d}R^{'}
\;  R^{'2} \;  e^{\Phi_2(R^{'})}
+\int_{R}^{R_{max}}{\rm d}R^{'} \; R^{'} \; e^{\Phi_2(R^{'})}
\right)\nonumber\\
\Phi_2(R)&=&a_2+4 \pi \, \frac{\eta_1}{\mu} \left(\frac{1}{R}\int_{0}^{R} {\rm
d}R^{'} \;  R^{'2} \; e^{\Phi_1(R^{'})}+\int_{R}^{R_{max}}{\rm d}R^{'}
\; R^{'} \; e^{\Phi_1(R^{'})} \right)\nonumber\\ 
&&+4 \pi \, \eta_2\left(\frac{1}{R}\int_{0}^{R} {\rm d}R^{'} \;  R^{'2} \; 
e^{\Phi_2(R^{'})}+\int_{R}^{R_{max}}{\rm d}R^{'} \; R^{'} \; e^{\Phi_2(R^{'})}
\right) \; . 
\end{eqnarray}
\noindent
We introduce the densities for the two kinds of particles at the
boundary ($R=R_{max}$)  
\begin {equation} \label{densper}
f_1 \equiv e^{\Phi_1(R_{max})}=\frac{\lambda^2}{3 \eta_1^R} \;
e^{\chi_1\left(\lambda\right)}\quad , \quad
f_2\equiv e^{\Phi_2(R_{max})}=\frac{\lambda^2\; e^{C}}{3 \, \eta_2^R} \;
e^{\frac{\chi_1\left(\lambda\right)}{\mu}} \; ,
\end{equation}
where we used eqs.(\ref{phi}) and (\ref{dilatation}).
Notice that $ \lambda $ and $ C $ are functions of $ \eta_1^R $ and $
\eta_2^R $ as explained by the end of sec. 3.1. Hence, $f_1$ and $f_2$ are
functions of $ \eta_1^R $ and $ \eta_2^R $.

We find for $R=R_{max}$ the following expressions for the Lagrange
multipliers using the normalization of the densities 
(\ref {normalisation}) 
\begin{equation} \label{Lagrange}
a_1=\ln f_1-\eta_1^R-\mu \, \eta_2^R  \quad , \quad
a_2=\ln f_2-\frac{1}{\mu} \,\eta_1^R-\eta_2^R \; .
\end{equation}
\noindent
Inserting these expressions (\ref {Lagrange}) into the free
energy (\ref {free1}), we find 
\begin {eqnarray}
\label{free2}
\frac{F-F_0}{T}&=&\frac{N_1}{2}
\left[\ln f_1 -\eta_1^R-\mu \, \eta_2^R \right]+
\frac{N_2}{2}
\left[\ln f_2 -\frac{1}{\mu}\, \eta_1^R-\eta_2^R \right]\\
&&+\frac{N_1}{2}
\int {\rm d}^3{\bf x}\;\Phi_1({\bf x})\;e^{\Phi_1({\bf x})}
+\frac{N_2}{2}
\int {\rm d}^3{\bf x}\;\Phi_2({\bf x})\;e^{\Phi_2({\bf x})} \; . \nonumber
\end{eqnarray}
\noindent
We compute the integrals in the second line in appendix A and we find
$$ 
\frac{F-F_0}{T}=N_1
[\ln f_1 -\eta_1^R-\mu \; \eta_2^R +3(1-f_1)]+
N_2 [\ln f_2 -\frac{1}{\mu} \; \eta_1^R-\eta_2^R +3(1-f_2)]\;.
$$ 
\noindent
The free energy as well as the other physical quantities are functions
of $\eta_1^R$ and $\eta_2^R$. The parameters $\eta_1^R$ and $\eta_2^R$
are linked by the relation, 
$$ 
\eta_2^R= \frac{m_2^2 \; N_2}{m_1^2 \; N_1}\; \eta_1^R \; .
$$ 
Hence, for fixed ratios $\frac{N_1}{N_2}$ and $ \frac{m_2}{m_1} $, 
the physical quantities depend only on $\eta_1^R$ or on $\eta_2^R$. In
that case it is simpler to see the physical quantities as functions of
$\eta_2^R$ or $\eta_1^R$ on a two dimensional plot than to watch the three
dimensional  surfaces for the physical quantities as 
functions of $\eta_2^R$ {\bf and} $\eta_1^R$. 

We plot in fig.\ref{freeen} the free energy as a function of
$\eta_2^R$ for fixed $\frac{\eta_1^R}{\eta_2^R}=\frac{16}{3}$.
In the limit where the  particles of mass $m_1$ dominate ($N_1 \gg N_2$) the
free energy becomes
$$ 
\frac{F}{T}\buildrel{N_1 \gg N_2}\over=N_1[\ln f_1 -\eta_1^R+3(1-f_1)]\;.
$$ 
\noindent
We recognize here the free energy of a self-gravitating gas with 
$N_1$ particles of mass $m_1$ (see ref.\cite{1sg}).

\subsection{Energy}

We compute here the gravitational energy of the gas.
The density of gravitational energy is 
$$
\epsilon_{P}({\bf x})=\frac{1}{2}\left(\frac{m_1 N_1 \rho_1({\bf x})}{V}+
\frac{m_2 N_2 \rho_2({\bf x})}{V} \right) U({\bf x}) 
$$
 \noindent
where $U$ is the gravitational potential (\ref {potentiel}).
Hence,
\begin{equation}   \label{densengrav2}
\epsilon_{P}({\bf x})=-\frac{N_1 T}{2 V} \; e^{\Phi_1({\bf x})} \;
[\Phi_1({\bf x})-a_1] -\frac{N_2 T}{2 V} \;  e^{\Phi_2({\bf x})} \;
[\Phi_2({\bf x})-a_2]\;. 
\end{equation}
\noindent
Using  the expressions for the Lagrange multipliers (\ref{Lagrange}), 
we obtain for the gravitational energy density
in the spherically symmetric case
\begin{eqnarray}   \label{densen2}
\epsilon_{P}({\bf x})&=&\frac{N_1 T}{2 V}  \; \rho_1(R)
\left[\ln\left(\frac{\rho_1(R_{max})}{\rho_1(R)}\right)-\eta_1^R-
\mu\eta_2^R\right]\nonumber\\ 
&&+ \frac{N_2 T}{2 V} \;  \rho_2(R)
\left[\ln\left(\frac{\rho_2(R_{max})}{\rho_2(R)}\right)
-\frac{1}{\mu}\eta_1^R-\eta_2^R \right] \; .
\end{eqnarray}
\noindent
Integrating the energy density (\ref{densengrav2}), with the help of
eqs. (\ref {normalisation}) and (\ref {Lagrange})  we obtain
\begin{eqnarray}
\label{interen}
E_P&=&\frac{N_1 \, T}{2}\left[\ln{f_1}-\eta_1^R-\mu \; \eta_2^R\right]
+\frac{N_2 \, T}{2}\left[\ln{f_2}-\frac{1}{\mu} \; \eta_1^R-
\eta_2^R\right]\nonumber \\
&-&\frac{N_1 \, T }{2}
\int {\rm d}^3{\bf x}\;\Phi_1({\bf x})\;e^{\Phi_1({\bf x})}
-\frac{N_2 \, T}{2}
\int {\rm d}^3{\bf x}\;\Phi_2({\bf x})\;e^{\Phi_2({\bf x})} \; .
\end{eqnarray}
\noindent
The integrals in the second line are computed in appendix A yielding for
the gravitational energy 
\begin{equation}   
\label{energy}
E_P=3 T\left[N_1(f_1-1)+N_2(f_2-1)\right]\;.
\end{equation}

\subsection{Entropy}

Using the gravitational energy (\ref{energy}) and the free energy
(\ref {free2})  we obtain for the entropy 
$$ 
S=S_0+N_1 [6(f_1-1)-\ln{f_1}+ \eta_1^R+\mu \; \eta_2^R ]
+N_2 [6(f_2-1)-\ln{f_2}+ \frac{1}{\mu} \; \eta_1^R+\eta_2^R ]
$$ 
\noindent
where $S_0$ is the entropy of the perfect gas.
$$ 
S_0=N_1 \left(\ln{\left[\frac{V}{N_1} \left(\frac{m_1 T}{2 \pi}\right)^
{\frac{3}{2}}\right]}+\frac{5}{2}\right)
+N_2 \left(\ln{\left[\frac{V}{N_2} \left(\frac{m_2 T}{2 \pi}\right)^
{\frac{3}{2}}\right]}+\frac{5}{2}\right)\;.
$$ 
We plot in Fig. \ref{entropy} the entropy $ S $ as a function of $ \eta_2^R
$ for $\frac{\eta_1^R}{\eta_2^R}=\frac{16}{3}$.

\subsection{Local pressure}

Since the system is non-homogeneous the local pressure is not uniform.
The density of gravitational force is
$$ 
{\bf F}({\bf r})=-\frac{m_1 \; N_1 \; \rho_1({\bf r})+
m_2 \; N_2 \; \rho_2({\bf r})}{V} \; {\bf grad}[U({\bf r})]
$$ 
\noindent
where U({\bf r}) is the gravitational potential (\ref {potentiel}).

\noindent
We obtain for the force at the point $ {\bf r} $ using eqs.(\ref{phi})
and (\ref{potentiel}),
$$ 
{\bf F}({\bf r})=\frac{N_1 T}{V} \; {\bf grad}(e^{\Phi_1({\bf r})})
+\frac{N_2 T}{V} \; {\bf grad}(e^{\Phi_2({\bf r})})\;.
$$ 
\noindent
The link between the density of force and the pressure is
$$ 
 {\bf F}({\bf r})={\bf grad}[P({\bf r})]\;.
$$ 
\noindent
Using eq.(\ref{phi}),  the local pressure is given by,
\begin {equation} \label{pessure}
P({\bf r})=\frac{N_1 T}{V} \; \rho_1({\bf r}) +
\frac{N_2 T}{V} \; \rho_2({\bf r}) \;.
\end{equation} 
\noindent
This is the {\bf local} equation of state for the self-gravitating gas 
with two kinds of particles. We see that it locally coincides with the
equation of state of a perfect gas. Since the gas is inhomogeneous,
the pressure acting on any finite volume will {\bf not} obey the ideal
equation of state.

\begin{figure}[htbp]
\rotatebox{-90}{\epsfig{file=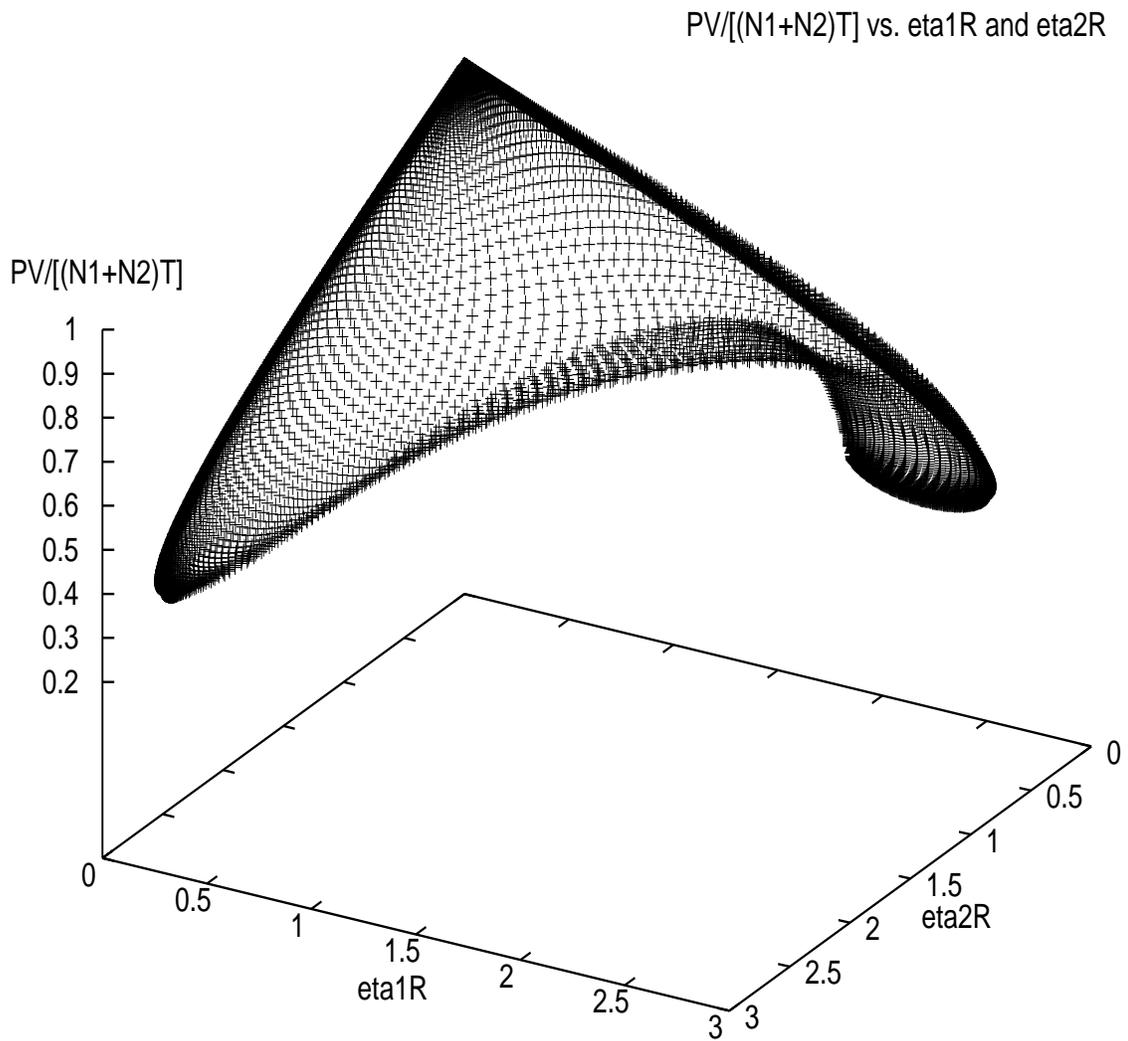,width=16cm,height=18cm}}
\caption{Equation of state: $\frac{PV}{(N_1+N_2)T}$  
versus $\eta_1^R$ and  $\eta_2^R$. We choose
$\frac{m_1}{m_2}=4$.}
\label{pressure3d}
\end{figure}

For a point ${\bf r}$ at the boundary, using eqs.(\ref{phi}),
(\ref{densper}) and (\ref{pessure}) yields the external pressure as a
function of $\eta_1^R$ and $\eta_2^R$ as,
\begin {equation} \label{pessureext}
P=\frac{N_1 T}{V} f_1+\frac{N_2 T}{V} f_2 \; ,
\end{equation}
\noindent
where $f_1$ and $f_2$ are defined by eq.(\ref{densper}).
This is the equation of state of the gas as a whole that we plot in
fig.\ref{pressure3d}. 

Combining eqs.(\ref{energy}) and (\ref{pessureext})
yields the virial theorem
$$ 
\frac{PV}{T}=N_1+N_2+\frac{E_P}{3 T}\;.
$$ 

\begin{figure}[h]
\rotatebox{-90}{\epsfig{file=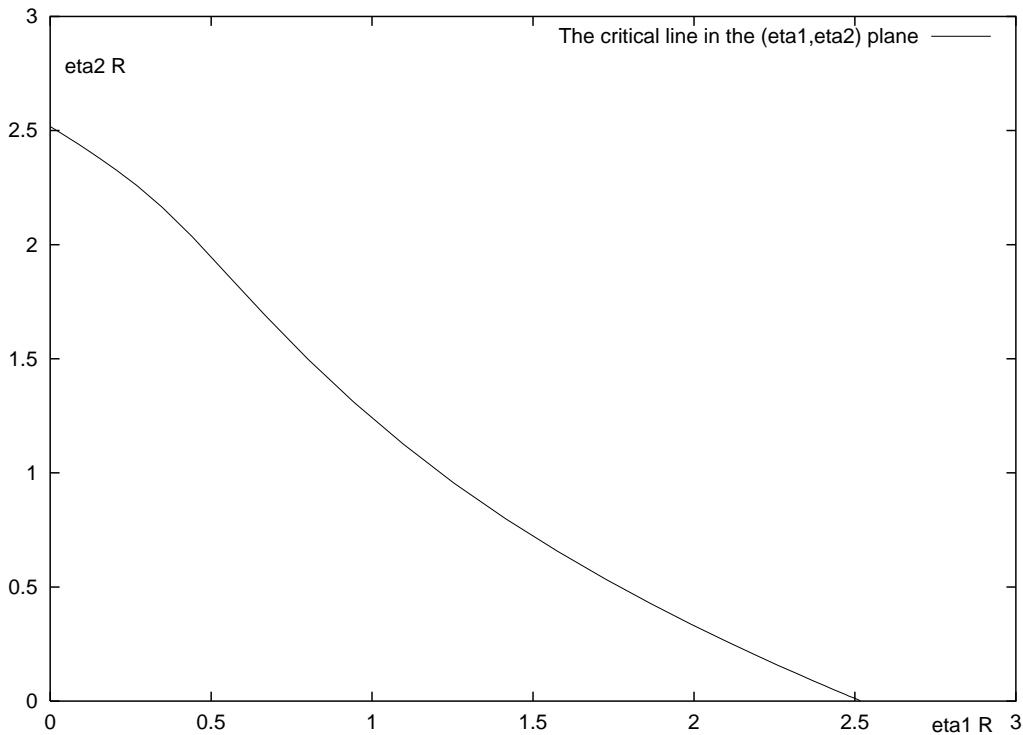,width=10cm}}
\caption{The critical line in the $(\eta_1^R,\eta_2^R)$ plane.
The mean field approximation  is valid in the region below the critical line}
\label{critline}
\end{figure}

\begin{figure}[htbp]
\rotatebox{-90}{\epsfig{file=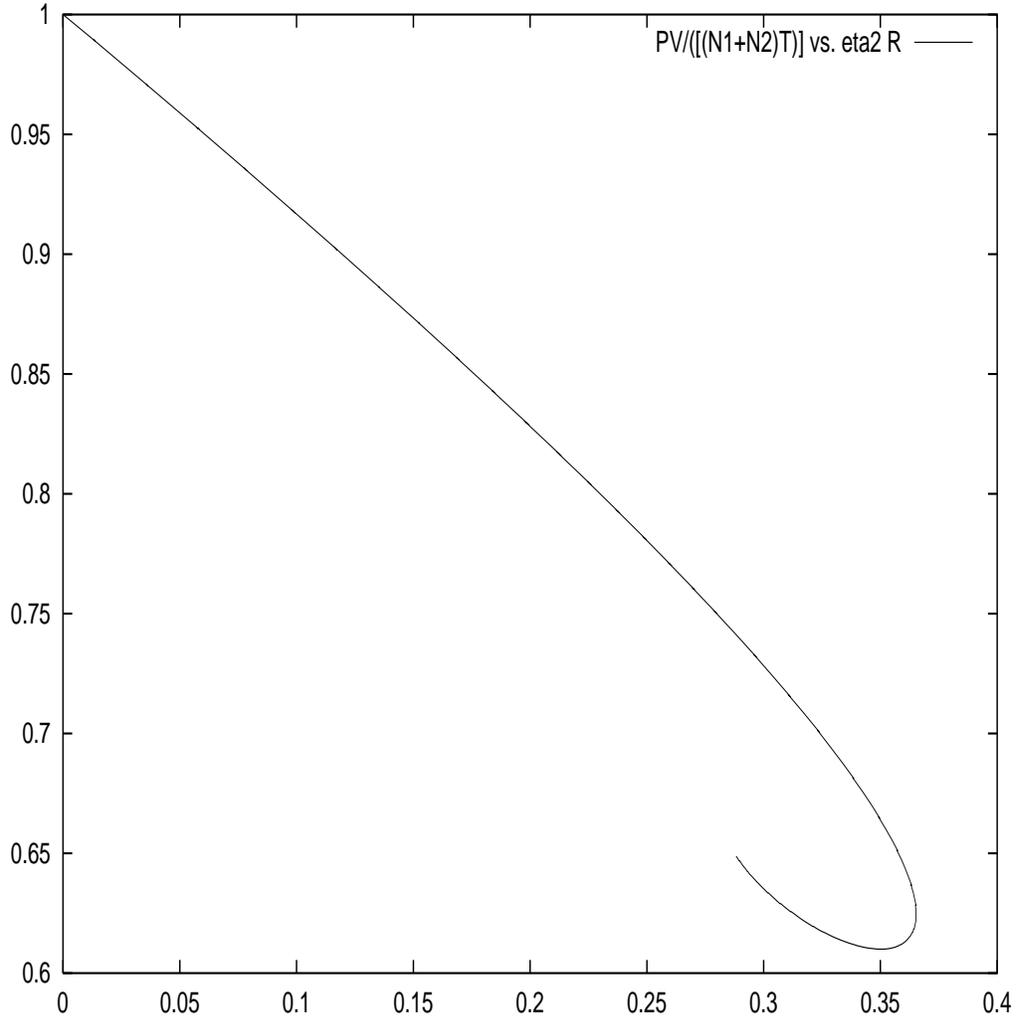,width=14cm,height=14cm}}
\caption{Equation of state $\frac{PV}{(N_1+N_2)T}$  as a function of 
$\eta_2^R$ for 
$\frac{m_1}{m_2}=4, \; \frac{N_1}{N_2}=\frac{1}{3}$ and therefore  
$\frac{\eta_1^R}{\eta_2^R}=\frac{16}{3}$.} 
\label{pressuresec}
\end{figure}

\begin{figure}[h]
\rotatebox{-90}{\epsfig{file=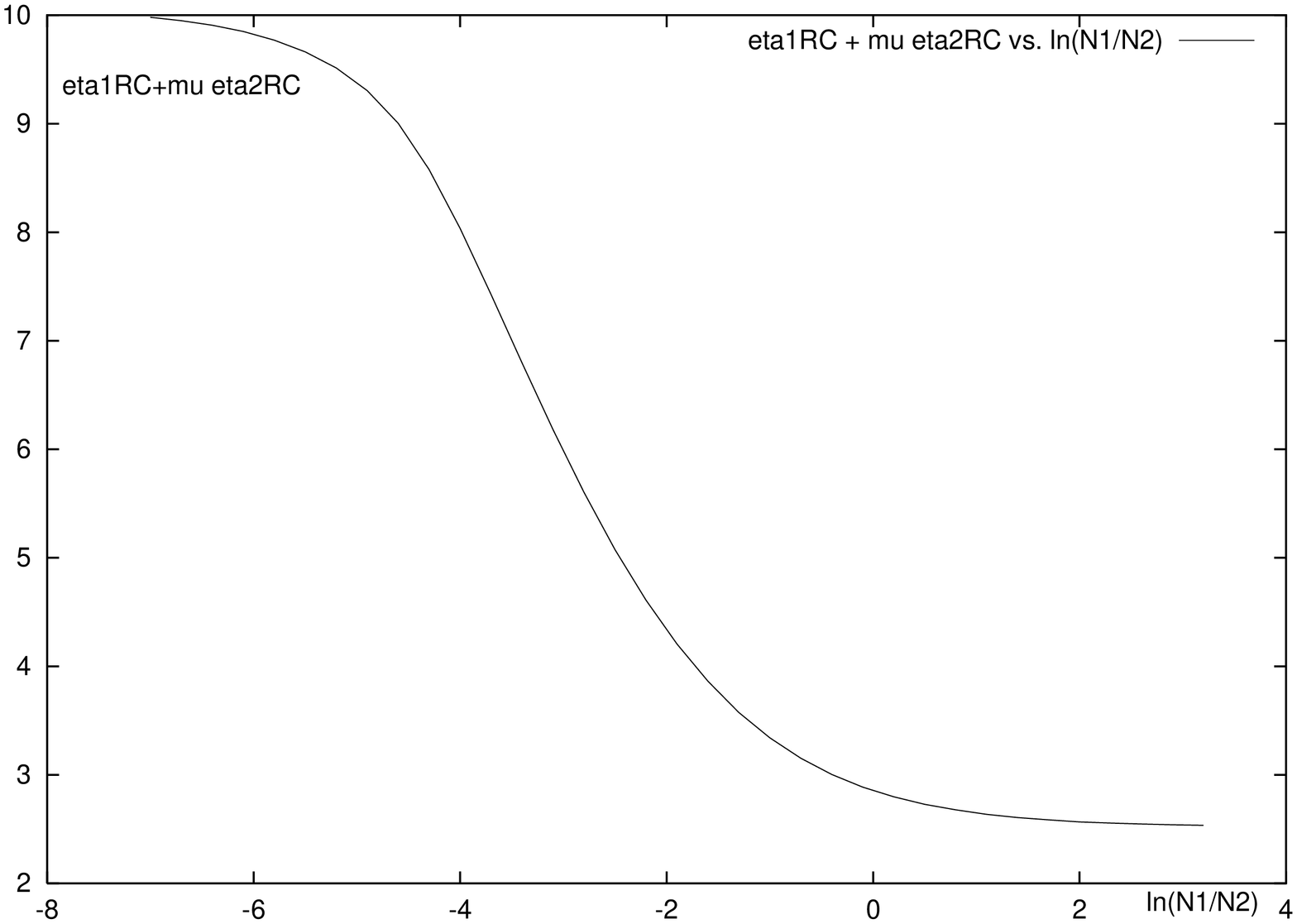,width=14cm,height=14cm}}
\caption{Critical value of the physical parameter 
$\eta_1^{RC}+\mu \; \eta_2^R$ versus
$\ln{\frac{N_1}{N_2}}$ for $\frac{m_1}{m_2}=4$.}
\label{etabc}
\end{figure}

\subsection{Physical behaviour of the system}

This self-gravitating system formed by two kinds of particles can be in two
phases: gaseous and condensed. The first one corresponds to $ \eta_1 $
and $ \eta_2 $ between the origin and their collapse values. In the
gaseous phase the free energy has a minimum for the pair of densities
$(\rho_1,\rho_2)$, solutions of the saddle point equations (\ref
{sadeq}). This pair of densities is the most probable distribution
which become absolutely certain in the thermodynamic limit. All
thermodynamic quantities follows from this  pair of densities
$(\rho_1,\rho_2)$.  

In the condensed phase, $(\rho_1,\rho_2)$ from mean field does not
describe the particle distribution and the mean field approach fails
to describe the condensed phase. It may be studied by Monte-Carlo
methods as in ref.\cite {1sg}.

\begin{figure}[h]
\rotatebox{-90}{\epsfig{file=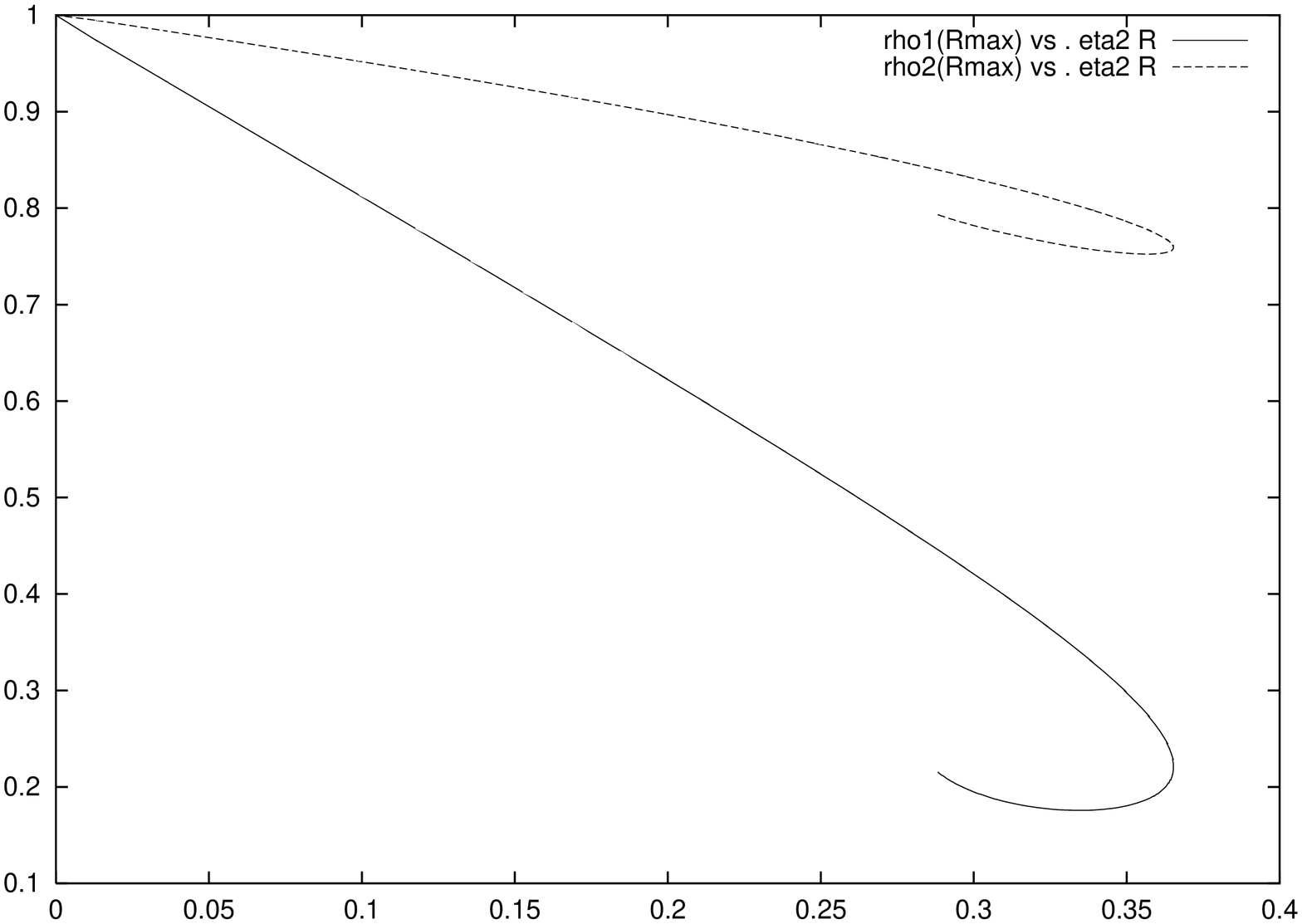,width=14cm,height=14cm}}
\caption{Density of particles of mass $m_1 \; [ \rho_1(R_{max}) ]$
and density of particles of mass $m_2\; [ \rho_2(R_{max}) ] $ 
at the boundary versus $\eta_2^R$ where the mass ratio  
is $\frac{m_1}{m_2}=4$, the number of particles ratio is
$\frac{N_1}{N_2}=\frac{1}{3}$ and then
$\frac{\eta_1^R}{\eta_2^R}=\frac{16}{3}$.}
\label{densbound}
\end{figure}

\begin{figure}[h]
\rotatebox{-90}{\epsfig{file=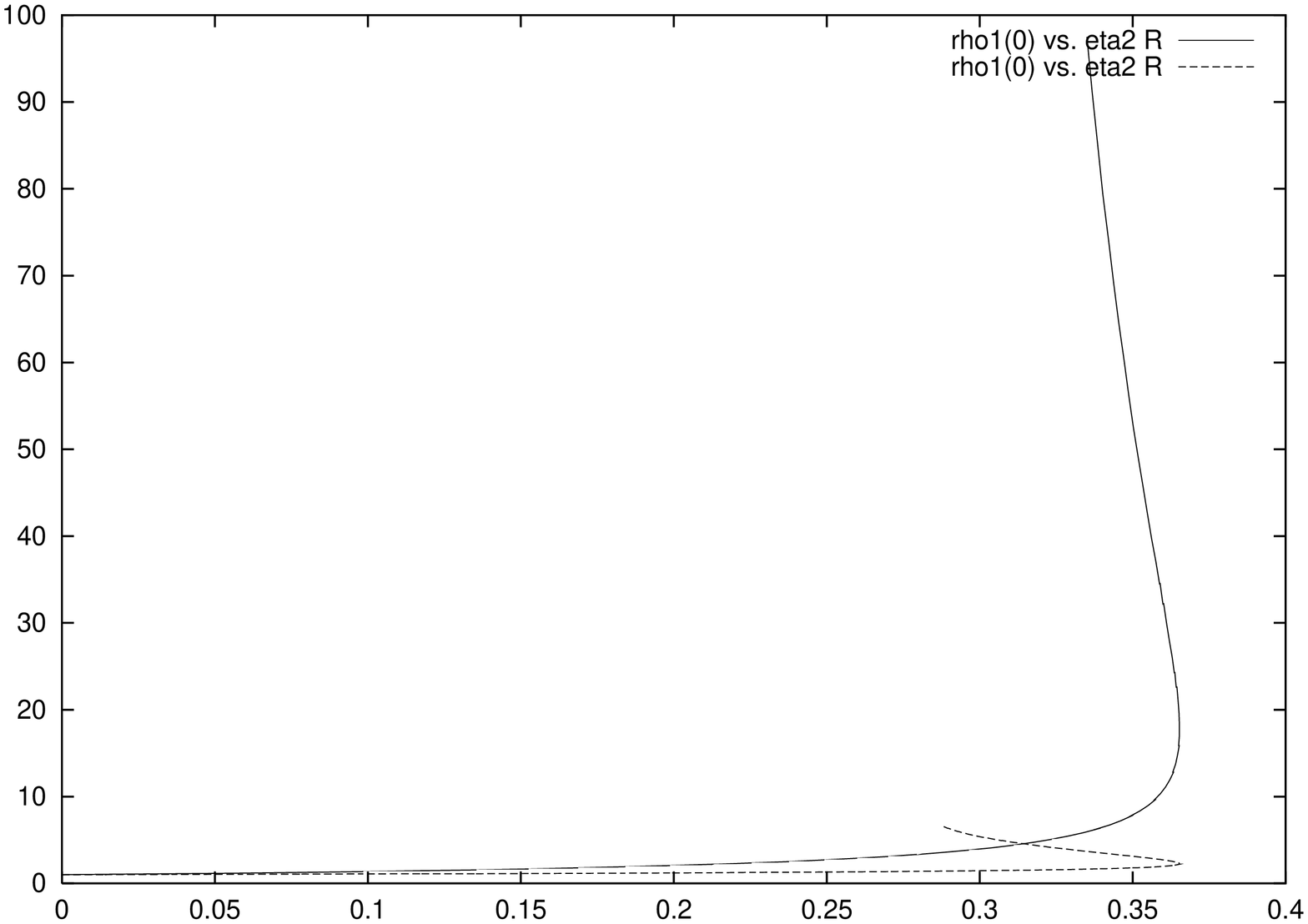,width=14cm,height=14cm}}
\caption{Density of particles of mass $ m_1 \; [ \rho_1(0) ] $ and
density of particles of mass $ m_2(0) \; [ \rho_2(0) ] $ at the origin
versus $\eta_2^R$ where the 
mass ratio is $\frac{m_1}{m_2}=4$, the number of particles ratio is
$\frac{N_1}{N_2}=\frac{1}{3}$ and then
$\frac{\eta_1^R}{\eta_2^R}=\frac{16}{3}$.}
\label{densor}
\end{figure}

When the physical parameters $\eta_1^R=\eta_2^R=0$ we retrieve the
perfect gas. When $\eta_1^R$ and $\eta_2^R$ increase, the gas becomes
denser at the center of the sphere ($R=0$) and less dense at the
boundary $(R=R_{max})$ because of gravitational attraction [see
Figs. \ref{densbound}-\ref{densor}]. This effect is more acute for the
heavier particles showing that more massive particles diffuse to the
denser regions. 

The equation of state is depicted in fig.\ref{pressure3d}. In the
ideal gas limit, $\eta_1^R=\eta_2^R=0,\; PV=(N_1+N_2)T $.  In the case
$\eta_1^R=0$ (gas of particles of mass $m_2$) and in the case
$\eta_2^R=0$ (gas of particles of mass $m_1$) we recover the equation
of state of the self-gravitating gas with one kind of particles
\cite{1sg}.

We call critical values $\eta_1^{RC}$ and $\eta_2^{RC}$ the points
where $ PV/[(N_1+N_2) \, T] $ exhibits a vertical slope. $\eta_1^{RC}$
and $\eta_2^{RC}$ define a critical line in the $(\eta_1^R, \eta_2^R)$
plane.  Namely, for each value of $N_1/N_2$ we have a different pair of
critical points $\eta_1^{RC}$ and $\eta_2^{RC}$. We plot the critical
line in the $(\eta_1^R, \eta_2^R)$ plane in fig.\ref{critline}.

The surface pressure has a rim on the canonical critical line [see
fig. \ref{pressure3d}]. The projection of this rim on the $(\eta_1^R,
\eta_2^R)$ plane yields the critical line plotted in
fig.\ref{critline}.  For a fixed $ N_1/N_2 $ we get a section of the
equation of state surface depicted in fig.\ref{pressuresec}.  This
section turns to have a form analogous to the equation of state for
the self-gravitating gas with one kind of particles \cite{1sg}.

By analogy with the gas with one kind of particles, we expect that the
gas collapses in the condensed phase for values of $\eta_1^R$ and
$\eta_2^R$ slightly below $\eta_1^{RC}$ and $\eta_2^{RC}$ where the
saddle point approximation breaks down.

The physical quantities exhibit a square root Riemann sheet structure
as functions of $\eta_1^R$ and $\eta_2^R$. The branch points are on
the critical line for the canonical ensemble.  The lower branch [see
Figs. \ref{pressure3d} and \ref{pressuresec} for the equation of state
and figs.  \ref{densbound}-\ref{densor} for the densities] describe a
phase absent in the canonical ensemble. Such phase is realized and is
stable in the microcanonical ensemble as it was the case for the gas
with one kind of particles\cite{1sg}.

\begin{figure}[h]
\rotatebox{-90}{\epsfig{file=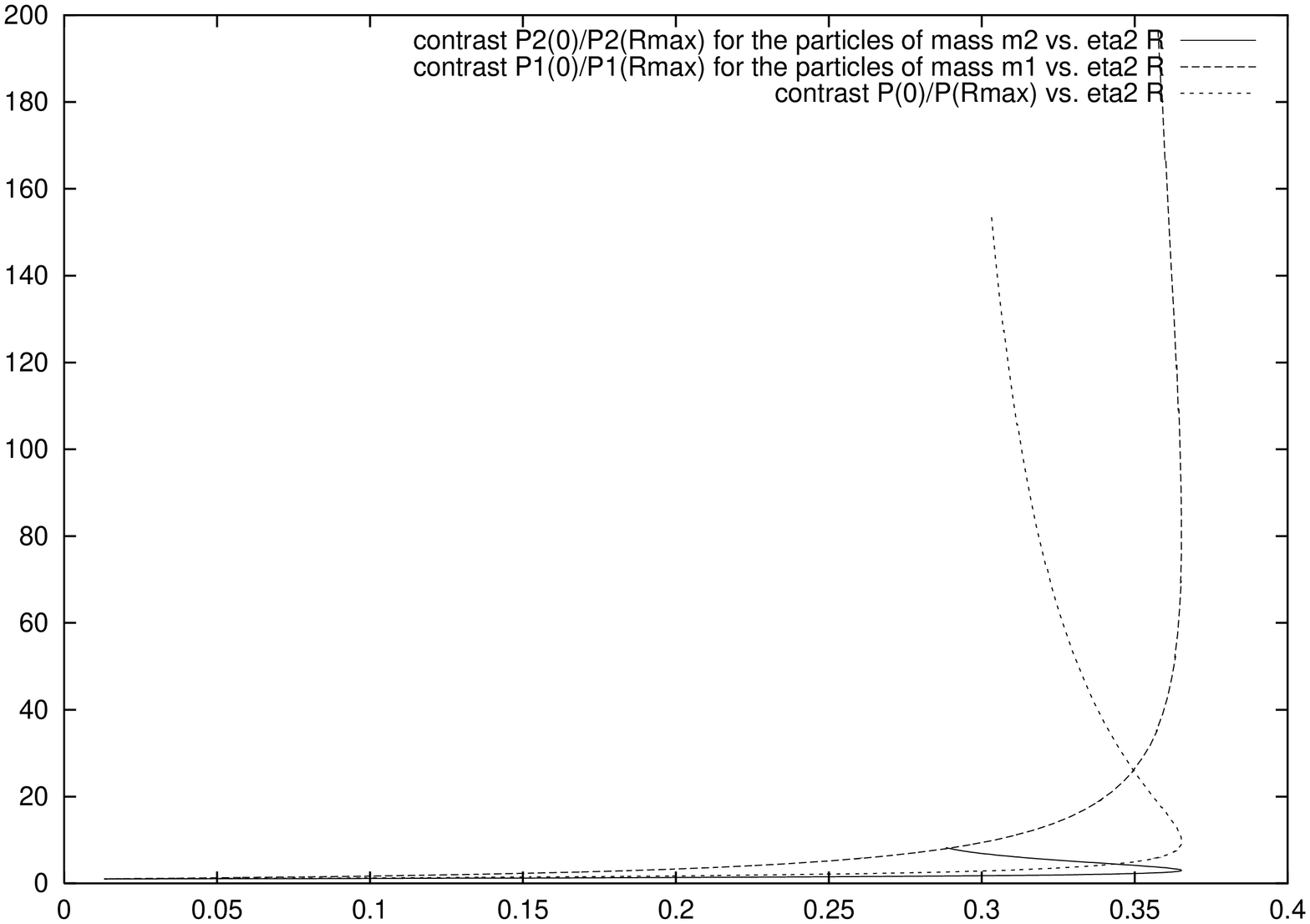,width=14cm,height=14cm}}
\caption{Contrast $P(0)/P(R_{max})$, partial contrast
$P_1(0)/P_1(R_{max})$ for the particles of mass $m_1$ and partial
contrast $P_2(0)/P_2(R_{max})$ for the particles of mass $m_2$ versus
$\eta_2^R$ where the mass ratio is $\frac{m_1}{m_2}=4$, the number of
particles ratio is $\frac{N_1}{N_2}=\frac{1}{3}$ and then
$\frac{\eta_1^R}{\eta_2^R}=\frac{16}{3}$. We have at the critical
point $P_1(0)/P_1(R_{max}) \simeq 81. > P(0)/P(R_{max})\simeq 9.9 >
P_2(0)/P_2(R_{max})\simeq 3.0 $. }
\label{contrast}
\end{figure}

We plot in Fig.\ref{etabc} the value of $ \eta_1^{R}+\mu \, \eta_2^{R}
$ at the critical line as a function of the number of particles $ N_1
/N_2 $. This quantity is proportional to the total mass of the gas $
m_1 \, N_1+m_2\, N_2 $.  This critical parameter $\eta_1^{RC}+\mu \;
\eta_2^{RC}$ interpolates between the two limiting cases ($N_1 \gg
N_2$ and $N_1 \ll N_2$) where one kind of particles dominate over the
others.  When $ N_1 \gg N_2 $ the particles of mass $m_1$ dominate and
$\eta_1^{RC} \gg \eta_2^{RC}$. In this limiting case, $\eta_1^{RC}+\mu
\; \eta_2^{RC} \to \eta^{RC} = 2.518\ldots$ which is the critical value
for the self gravitating gas with one kind of particles associated to
the Jeans instability\cite{hidro,1sg}. When $ N_2 
\gg N_1 $ the particles of mass $m_2$ dominate and
$\eta_2^{RC}\gg\eta_1^{RC}$. In this limiting case, $\eta_1^{RC}+\mu
\; \eta_2^{RC} \to \mu \, \eta^{RC} = 10.04\ldots $ for the mass ratio
$\mu=4$ corresponding to a mixture of hydrogen and helium.

\bigskip

From eq.(\ref{pessure}) we see that the partial pressures of the
particles with mass $m_1$ and $m_2$ are given by,
\begin{equation} \label{pres12}
P_1({\bf r})=\frac{N_1 T}{V} \; e^{\Phi_1({\bf r})} \quad , \quad
P_2({\bf r})=\frac{N_2 T}{V} \; e^{\Phi_2({\bf r})}\;.
\end{equation}
The pressure contrast is defined by the ratio of the pressure at the
center and the pressure at the boundary: $P(0)/P(R_{max})$
\cite{hidro}. We find from eqs.(\ref{limcond2}), (\ref{qui20}),
(\ref{dens}) and (\ref{pessure})
$$
\alpha \equiv \frac{P(0)}{P(R_{max})} = \frac{ 1 + \mu^2 \, e^{C}}
{e^{\chi_1(\lambda)} + \mu^2 \, e^{C + \chi_1(\lambda)/\mu}} \; .
$$
We extend this definition to each kind of particles and say that the
partial contrasts are given by $P_1(0)/P_1(R_{max})$ for the particles
of mass $m_1$ and $P_2(0)/P_2(R_{max})$ for the particles of mass
$m_2$.  We find from eqs.(\ref{limcond2}), (\ref{qui20}), (\ref{dens})
and (\ref{pres12}),
\begin{equation} \label{contras}
\alpha_1 \equiv \frac{P_1(0)}{P_1(R_{max})} = e^{-\chi_1(\lambda)}
\quad \mbox{and} \quad \alpha_2 \equiv \frac{P_2(0)}{P_2(R_{max})} =
e^{-\chi_1(\lambda)/\mu} \; .
\end{equation}
We plot the contrast and the partial contrasts in fig. \ref{contrast}.
We see that the contrast $\alpha$ takes here {\bf lower} values than
for a gas with one kind of particles. On the contrary, $\alpha_1$, the
partial contrast for the {\bf heavier} particles, takes {\bf higher
values} than the contrast for a gas with one kind of particles. This
is due to the fact that the overdensity of particles in the center is
more acute for the heavier ones as noticed above.

The particle density and pressure has its maximum at the origin and its
minimum at the boundary, as expected. However, their ratio (contrast)
is much {\bf larger} for the heavier particles than for the lighter
ones. They are related through [see eq.(\ref{contras})],
$$
\alpha_1 = [\alpha_2]^{\mu} 
$$
where $ \mu = m_1/m_2 $. Since $ \alpha_2 > 1 $, when $ m_1 > m_2 $, we can
get $ \alpha_1 \gg \alpha_2 $ [see fig.\ref{contrast}]. In particular,
$ \alpha_1 $ may be much larger than the contrast in the gas with one
kind of particles\cite{hidro,1sg}. In summary, this shows that the
heavier particles diffuse to the denser regions. 

\subsection{Scaling law}

We compute here the mass $M(R)$ inside a sphere of radius $R$ ($0 \leq
R \leq R_{max}$).

Using Gauss's theorem and recalling that $U$ is the gravitational
potential (\ref {potentiel}) we find that
$$
M(R)=-\frac{m_1 N_1}{\eta_1} \; R^2 \; \Phi_1^{'}(R) \; .
$$
Using eqs.(\ref {dilatation}) we obtain
$$ 
M(R) =-\frac{m_1 \, N_1 \,
\lambda}{\eta_1^R} \; \left(\frac{R}{R_{max}}\right)^2 \;
\chi_1^{'}\left(\lambda \; \frac{R}{R_{max}} \right) \;.
$$ 
\noindent
As for the self-gravitating gas where all particles have the same
mass, the mass $M(R)$ for the self-gravitating gas with two kinds of
particles follows approximately the scaling law
$$ 
M(R) \approx {\cal C} \; R^d \;.
$$ 
\noindent
This indicates a fractal distribution with Haussdorf dimension $d$.

$d$ decreases with $\eta_1^R$ and $\eta_2^R$ from the value $d=3$ for
the ideal homogeneous gas ($\eta_1^R=\eta_2^R=0$) till $ d\approx 1.6
$ in the canonical critical line.  The Haussdorf dimension keeps
decreasing beyond the canonical critical line in the stable phase of
the microcanonical ensemble.

\begin{figure}[h]
\rotatebox{-90}{\epsfig{file=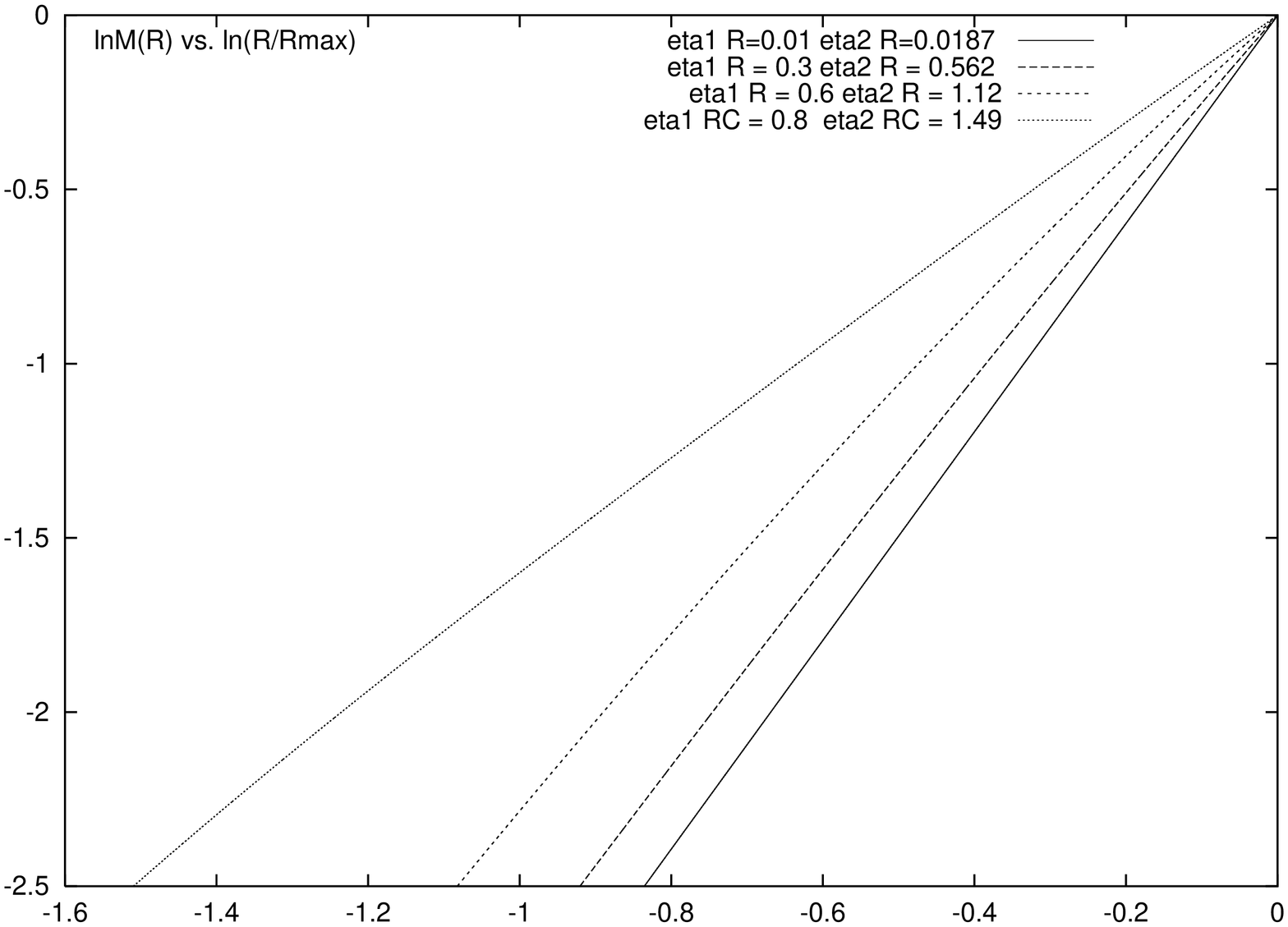,width=14cm,height=14cm}}
\caption{$\ln{M(R)}$ versus $\ln{\frac{R}{R_{max}}}$ for
$\frac{m_1}{m_2}=4, \; \frac{N_1}{N_2}=0.0334\ldots$ and then
$\frac{\eta_1^R} {\eta_2^R}=0.534\ldots$ for different values of
$\eta_1^R: \; \eta_1^R=0.01 , \; \eta_1^R=0.03 , \; \eta_1^R=0.06 $
and the canonical critical point $\eta_1^{RC}=0.8002\ldots $ }
\label{fractal2}
\end{figure}

\begin{figure}[h]
\rotatebox{-90}{\epsfig{file=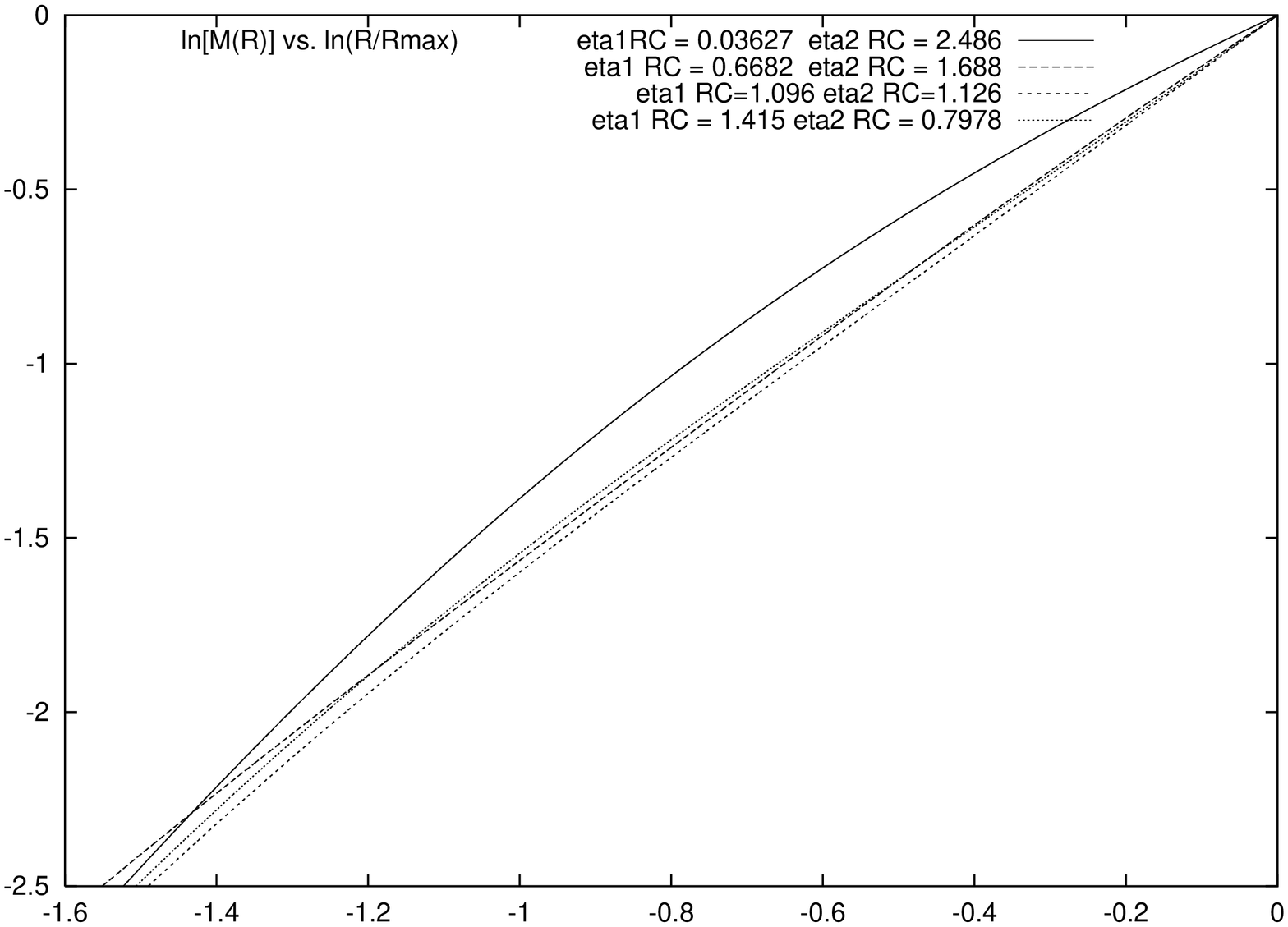,width=14cm,height=14cm}}
\caption{$\ln{M(R)}$ versus $\ln{\frac{R}{R_{max}}}$ for
$\frac{m_1}{m_2}=4$ on the critical canonical line for different
values of $\eta_1^R$ and $ \eta_2^R $: 
$ \; \eta_1^{RC}=0.03627 \ldots , \;\eta_2^{RC}=2.486\ldots \; ; 
\eta_1^{RC}=0.6682\ldots , \; \eta_2^{RC}=1.688\ldots \; ; 
\eta_1^{RC}=1.096\ldots , \; \eta_2^{RC}= 1.126\ldots $
 and $\eta_1^{RC}=1.415\ldots , \; \eta_2^{RC}=0.7978\ldots$
The Haussdorf dimension $d$ turns to be
independent of the composition of the gas taking the value $ d \approx
1.6 $. It coincides within the numerical accuracy with the Haussdorf
dimension for the gas with one kind of particles at the canonical
critical point\cite{1sg}. }
\label{fractalcrit}
\end{figure}

We plot in fig.\ref{fractal2} the mass $M(R)$ for $M(R)$ greater than
$10\%$ of the total mass of the gas for several values of $\eta_1^R$
choosing the ratio $\eta_1^R/ \eta_2^R $ to be $0.534\ldots$.  We
exclude the region $M(R)<0.1$ where the mass distribution is almost
uniform. This local uniformity is simply explained by the absence of
gravitational forces at the origin $R=0$ due to the spherical
symmetry.

\vspace{1cm}

\begin{tabular}{|l|l|l|l|}\hline
$ \eta_1^R $ & $\eta_2^R $ & $\hspace{0.5cm} d $ & $\hspace{0.5cm} C
$\\ \hline $ 0.01 $ & $ 0.0187 $ & \hspace{0.3cm} $ 2.99 $
\hspace{0.3cm} & \hspace{0.3cm} $ 1.00 $ \hspace{0.3cm} \\ \hline $
0.3 $ & $ 0.562 $ & \hspace{0.3cm} $ 2.72 $ \hspace{0.3cm} &
\hspace{0.3cm} $ 1.03 $ \hspace{0.3cm} \\ \hline $ 0.6 $ & $ 1.12 $ &
\hspace{0.3cm} $ 2.31 $ \hspace{0.3cm} & \hspace{0.3cm} $ 1.07 $
\hspace{0.3cm} \\ \hline $ \eta_1^{RC}=0.80... $ &
$\eta_2^{RC}=1.49...  $ & \hspace{0.3cm} $ 1.66 $ \hspace{0.3cm} &
\hspace{0.3cm} $ 1.03 $ \hspace{0.3cm} \\ \hline
\end{tabular}

 \bigskip

{TABLE 1. The fractal dimension $d$ and the proportionality
coefficient $ \cal C $ from a fit according to $ M(R) \approx {\cal C}
\; R^d$ for $\frac{m_1}{m_2}=4, \; \frac{N_1}{N_2}=0.0334 $ and then
$\frac{\eta_1^R} {\eta_2^R}=0.534$ and different values of $\eta_1^R$
and $\eta_2^R$.}
 
\vspace{1cm}
 
As shown in fig. \ref{fractalcrit}, the Haussdorf dimension at the
canonical critical line turns to be {\bf independent} of the ratio
$\eta_1^R / \eta_2^R $.  $d$ coincides within the numerical precision
with the Haussdorf dimension $ d \approx 1.6 $ at the canonical
critical point of the gas with one kind of particles \cite{1sg}.  This
indicates that the Haussdorf dimension at the canonical critical line
is an universal value, independent of the gas composition.

\begin{tabular}{|l|l|l|l|}\hline
$ \eta_1^{RC} $ & $\eta_2^{RC} $ & $\hspace{0.5cm} d $ &
$\hspace{0.5cm} C $\\ \hline $ 0.03627 $ & $ 0.0187 $ & \hspace{0.3cm}
$ 1.62 $ \hspace{0.3cm} & \hspace{0.3cm} $ 1.19 $ \hspace{0.3cm} \\
\hline $ 0.6682 $ & $ 1.688 $ & \hspace{0.3cm} $ 1.61 $ \hspace{0.3cm}
& \hspace{0.3cm} $ 1.04 $ \hspace{0.3cm} \\ \hline $ 1.096 $ & $ 1.126
$ & \hspace{0.3cm} $ 1.65 $ \hspace{0.3cm} & \hspace{0.3cm} $ 1.03 $
\hspace{0.3cm} \\ \hline $ 1.415 $ & $ 0.7978 $ & \hspace{0.3cm} $
1.62 $ \hspace{0.3cm} & \hspace{0.3cm} $ 1.04 $ \hspace{0.3cm} \\
\hline
\end{tabular}
 
\bigskip

{TABLE 2. The fractal dimension $d$ and the proportionality
coefficient $ {\cal C} $ from a fit according to $M(R) \approx{\cal C}
\; R^d $ on the critical canonical line for different values of
$\eta_1^{RC}$ and $\eta_2^{RC}$}

\subsection{Critical behaviour of the thermodynamic functions}

According to the behaviour of the free energy near the critical line
(see Fig. \ref{freeen}), we find that the first derivatives of the
free energy (energy, pressure) are continuous, while the second
derivatives (specific heats, compressibility) are discontinuous. Using
eq.(\ref {energy}) and the form of the functions $f_1$ and $f_2$ near
the critical line (see Fig. \ref{densbound}), we find that the energy
has two branches $E_{+}$ and $E_{-}$ which behave near the critical
point as,
\begin{equation} \label{enercrit}
E_{\pm}=\frac{3}{2}(N_1+N_2) T \pm D T\sqrt{\eta_2^{RC}-\eta_2^{R}}
\end{equation}
\noindent
where $D$ is a positive constant.  Deriving the energy
(\ref{enercrit}) with respect to $T$ and recalling eq.(\ref{etai}) for
$\eta_2$, we obtain the two branches of the specific heat at constant
volume
$$ 
C_{v_{\pm}}=\frac{3}{2}(N_1+N_2)
\pm \frac{D T}{2 \sqrt{\eta_2^{RC}-\eta_2^{R}}}\;.
$$ 
Here, $E_+$ and $C_{v+}$ stand for the gaseous stable phase  in the canonical
ensemble and $E_-$ and $C_{v-}$ correspond to a phase 
only realized in the microcanonical ensemble \cite{1sg}.
We see that $ E_+ = E_- $ at the critical point and hence 
$E$ is continuous at criticality  while $C_v$ exhibits there an
infinite discontinuity. (As is clear, negative values of $C_v$ cannot
be realized in the canonical ensemble\cite{llms}).

\section{A self-gravitating gas with $n$ kinds of particles}

The generalization of the treatment given in previous sections to a
self-gravitating gas with $n$ kinds of particles is  
straightforward. We give below the mean field equations and the more
relevant results. 

The relevant parameters of the gas are now
$$ 
\eta_i=\frac{G \, m_i^2 \, N_i}{L T}  \quad , \quad 1 \leq i \leq n, 
$$ 
\noindent
where $N_i$ is the number of particles of mass $m_i$. We assume that $
N_i/ L $ stay fixed while $ L \to \infty, \;   N_i\to
\infty,1 \leq i \leq n $. Therefore, the ratios $ N_i/N_j $ also stay fixed
for $ L \to \infty, \;   N_i\to \infty, 1 \leq i, \; j \leq n $.

The coupled mean-field integral equations for the densities of particles   
$\rho_i({\bf x})$ ($1 \leq i \leq n$) take the form
\begin {equation} \label{sadeq-n}
\ln{\rho_i({\bf x})}=a_i+m_i \; \sum_{j=1}^n \frac{\eta_j}{m_j}
\int\frac{{\rm d}^3{\bf y}}{|{\bf y}-{\bf x}|} \; \rho_j({\bf y}) 
\; \;  , 1 \leq i \leq n \;,
\end{equation}
\noindent
where Lagrange multipliers $a_1,a_2,...,a_n$ enforce the normalization
of the densities. 
\begin {equation}\label{normn}
\int {\rm d}^3{\bf x} \; \rho_i({\bf x}) = 1 \; ,  \;1 \leq i \leq n \; .
\end{equation} 
Eqs.(\ref{sadeq-n}) give for the gravitational potential
\begin{equation}   \label{potentieln}
U({\bf x})=-\frac{T}{m_1}\left[ \Phi_1({\bf x})-a_1 \right]
=-\frac{T}{m_2}\left[ \Phi_2({\bf x})-a_2\right] =...=
-\frac{T}{m_n}\left[\Phi_n({\bf x})-a_n\right]\;.   
\end{equation} 
\noindent
Setting, 
\begin {equation}\label{phin}
\rho_i({\bf x})=\exp\left[\Phi_i({\bf x})\right] \;  , \; 1 \leq i \leq n
\end{equation}
\noindent
and applying to eqs.(\ref{sadeq-n}) Laplace operator, we find the
partial differential equations
\begin {equation}\label{Poissonn}
\Delta \Phi_i({\bf x})+4 \pi \, m_i \sum_{j=1}^n  \frac{\eta_j}{m_j} \;
 e^{\Phi_j({\bf x})} =0 \;  \; , \; 1 \leq i,j\leq n .
\end{equation}
\noindent
Using eq.(\ref{potentieln}), we reduce eqs.(\ref{Poissonn}) to a single
equation
$$ 
\Delta \Phi_1({\bf x})+4 \pi \, m_1\sum_{j=1}^n \frac{\eta_j}{m_j} \;
e^{a_j-\frac{m_j}{m_1} a_1} \; e^{\frac{m_j}{m_1}\Phi_1({\bf x})} = 0
\; .
$$ 

\bigskip

Eqs. (\ref{Poissonn}) are scale covariant.
If $\Phi_1,\Phi_2,...,\Phi_n $ are 
solutions of eqs.(\ref {Poissonn}), then
$\Phi_{1 \lambda},\Phi_{2 \lambda},...,\Phi_{n \lambda}$
defined by
\begin {equation}\label{scaletransn}
\Phi_{i \lambda}({\bf x})=\Phi_i(\lambda {\bf x})+\ln{\lambda^2}
\; \; ,1 \leq i \leq n
\end{equation}
\noindent
are also solutions of eq.(\ref {Poissonn}). This property is due to
the scale behaviour of Newton's potential.

In the spherically symmetric case the mean field equations (\ref{Poissonn})
become ordinaries non-linear differential equations
\begin {equation}\label{Poissonradn}
\frac{{\rm d}^2 \Phi_i}{{\rm d}R^2}+\frac{2}{R} \frac {{\rm d} \Phi_i}{{\rm
d}R}
+4 \pi m_i \; \sum_{j=1}^n \; \frac{\eta_j}{m_j} \; e^{\Phi_j(R)} =0
\;  \; , \; 1 \leq i \leq n .
\end{equation}
\noindent
Using the scale covariance of the mean field eqs.(\ref {Poissonn})
by the transformation (\ref {scaletransn}), we can set
\begin {equation}\label{dilatationn}
\Phi_i(R)=\chi_i\left(\lambda \frac{R}{R_{max}}\right)
+\ln\left(\frac{\lambda^2}{3 \eta_i^R}\right)
\;  \; , \; 1 \leq i \leq n
\end{equation}
\noindent
with new parameters $ \eta_i^R=\frac{\eta_i}{R_{max}} , \;  1 \leq i \leq n$.
In this way, the mean field eqs.(\ref {Poissonradn}) become
a reduced system of the form
\begin {equation}\label{Poissonchin}
\chi_i^{''}(\lambda)+\frac{2}{\lambda} \; \chi_i^{'}(\lambda)+m_i
\sum_{j=1}^n \; \frac{e^{\chi_j(\lambda)}}{m_j} = 0
\;  \; , \; 1 \leq i \leq n.
\end{equation}
\noindent
Let us find the boundary conditions for these  equations.
In order to have a regular solution at the origin we impose
$$ 
\chi_i^{'}(0)=0
\;  \; , \; 1 \leq i \leq n \; .
$$ 
\noindent
The system (\ref{Poissonchin}) is invariant under the transformation
$$ 
\lambda \to \lambda \; e^{\alpha},\; \chi_i \to \chi_i-2 \alpha
\;  \; , \; 1 \leq i \leq n\; .
$$ 
\noindent
Hence, we can choose
$$ 
\chi_1(0)=0 
$$ 
\noindent
without loosing generality. As in the case of two kinds of particles, 
the remaining boundary conditions $\chi_2(0),...,
\chi_n(0) $ are not independent from $ \eta_1,\eta_2,...,\eta_n$.
The normalization (\ref{normn})
of the densities of the $n$ kinds of particles  $\rho_1, \; \rho_2,\ldots,
\rho_n$ has to be imposed. We obtain from eqs.(\ref{phin}) and
(\ref{dilatationn}), 
\begin {equation}\label{etai-normn}
\eta_i^R=\frac{1}{\lambda}
\int_{0}^{\lambda} {\rm d}x \; x^2 \; e^{\chi_i(x)}
\;  \; , \; 1 \leq i \leq n.
\end{equation}
\noindent
Using the reduced mean field equations (\ref {Poissonchin}), it is
straightforward to show  from eq.(\ref{etai-normn}) that
$$ 
m_i \sum_{j=1}^n \; \frac{\eta_j^R}{m_j}=-\lambda \; \chi_i^{'}(\lambda)
\;  \; , \; 1 \leq i \leq n.
$$ 
\noindent
Hence,
\begin {equation}\label{champn}
\chi_j^{'}(\lambda)=\frac{m_j}{m_i} \; \chi_i^{'}(\lambda) \;
\;  \; , \; 1 \leq i,j \leq n.
\end{equation}
\noindent
From eq.(\ref {champn})  we introduce new $\lambda$-independent parameters
$$ 
C_i=\chi_i(\lambda)-\frac{m_i}{m_1} \; \chi_1(\lambda)
\;  \; , \; 2 \leq i \leq n.
$$ 
\noindent
These new parameters are  only function of
$\eta_1^R, \; \eta_2^R,..., \;\eta_n^R$. Notice that the boundary
conditions can be written as,
$$
\chi_2(0)= C_2, \quad \ldots \quad ,\quad \chi_n(0)=C_n \; .
$$
The reduced mean field equations (\ref {Poissonchin}) become a single
ordinary differential equation with its  coefficients depending on 
the parameters $C_2 ,...,C_n$ ,
\begin {equation} \label{eqchi1n}
\chi_1^{''}(\lambda)+\frac{2}{\lambda}\chi_1^{'}(\lambda)
+m_1 \sum_{i=1}^n \; \frac{e^{C_i}}{m_i} \;
e^{\frac{m_i}{m_1} \; \chi_1(\lambda)}=0
\end{equation}
\noindent
and  the boundary  conditions $\chi_1(0)=0 ,\;\chi_1^{'}(0)=0$. [Here,
$C_1 \equiv 0$]. 

Using eqs.(\ref{phin}) and (\ref {dilatationn})
we can express the densities of the $n$  kinds of particles
in terms of the solution of eq.(\ref {eqchi1n})
$$ 
\rho_i(R)=\frac{\lambda^2 \; e^{C_i} }{3 \eta_i^R} \;
e^{\frac{m_i}{m_1}\chi_1(\lambda \frac{R}{R_{max}} )}\quad , 
 \quad 1 \leq i \leq n \;\;.
$$ 
The thermodynamic functions are
expressed as functions of the density of particles at the boundary
$f_1,f_2,...,f_n$ depending on $\eta_1^R, \; \eta_2^R,..., \;\eta_n^R$. 
That is,
$$
f_i=\frac{\lambda^2 \; e^{C_i} }{3 \eta_i^R} \;
e^{\frac{m_i}{m_1}\chi_1(\lambda)}\quad , 
 \quad 1 \leq i \leq n \;\;.
$$
We provide below the expressions for the free energy,  the gravitational 
energy,  the entropy and the equation of state. 
\begin{eqnarray}
\frac{F-F_0}{T}&=&\sum_{i=1}^n N_i \; [\; \ln{f_i}-\sum_{j=1}^n
\frac{m_i}{m_j}  \;\eta_j^R +3(1-f_i)\;] \cr \cr 
\label{energyn}
E_P&=&3 T \sum_{i=1}^n N_i \; (f_i-1) \\ \cr  
S&=&S_0+\sum_{i=1}^n N_i \; [\;6(f_i-1)-\ln{f_i}+\sum_{j=1}^n \frac{m_i}{m_j}
\; \eta_j^R\;] \label{pessureextn}\cr \cr
\frac{PV}{T}&=&\sum_{i=1}^n N_i \; f_i
\end{eqnarray}
\noindent
where 
$$ 
F_0=-\sum_{i=1}^n N_i \; T\; \ln{\left[\frac{e V}{N_i} \left(\frac{m_i T}{2
\pi}\right)^{\frac{3}{2}}\right]}
$$ 
\noindent
is the free energy and
$$ 
S_0=\sum_{i=1}^n N_i \left(\ln{\left[\frac{V}{N_i} \left(\frac{m_i
T}{2 \pi}\right)^{\frac{3}{2}}\right]}+\frac{5}{2}\right) 
$$ 
\noindent
is the entropy of the perfect gas with masses  $m_1, \; m_2,..., \; m_n$.
Combining eqs.(\ref{energyn}) and (\ref{pessureextn}) yields the
virial theorem
$$
\frac{PV}{T}=\sum_{i=1}^n N_i+\frac{E_P}{3 T}\;.
$$ 

\section{Conclusions}

The self-gravitating gas with two kinds of particle has analogous
qualitative properties to the self-gravitating gas with one  kind
of particles. Physical quantities like energy, free energy and entropy
turn to be the sum of a term proportional to $N_1$ plus another term
proportional to $N_2$ for large $N_1, \;N_2 $ and $V$ provided
$ N_1 / V^{\frac{1}{3}} $ and $ N_2 / V^{\frac{1}{3}} $ are kept fixed.
All physical quantities are expressed as functions of $\eta_1$ and
$\eta_2$. Instead of a critical line as for one kind of particles, we
have here a critical line in the $(\eta_1,\eta_2)$ plane for the
canonical ensemble . This line is associated to the Jeans instability.

The equation of state exhibits a rim on this
critical line [see fig.\ref{pressure3d}].
The thermodynamic functions exhibit a two-sheeted structure as
functions of  $\eta_1^R$ and $\eta_2^R$. The branch points are on the
critical line. The specific heat is discontinuous and diverges
there while the free energy is finite and continuous in the branch
line. 

The local pressure and the local densities of particles are related by
the same equation as in a perfect gas [see sec. 3.5]:
$$
P({\bf r})=\frac{N_1 T}{V} \; \rho_1({\bf r}) +
\frac{N_2 T}{V} \; \rho_2({\bf r}) \;.
$$
This can be explained by  the dilute character of the self-gravitating gas in
thermal equilibrium: $N/V \sim N^{-2} \to 0$ for $ N \to \infty $. This
dilution damps the effective interparticle interaction
and allows a free particle behaviour.

The particle distribution is inhomogeneous and scales
with $ R $ with a Haussdorf dimension $ d $.
The Haussdorf dimension $ d $ decreases for increasing $ \eta_1^R$ and $
\eta_2^R$. Its value on the critical line $ d =  1.6\ldots $ turns to
be independent of the ratio  $\eta_1^R/ \eta_2^R$ implying an
universal behaviour. $ d $ takes there the same value than for the
canonical critical point with one kind of particles \cite{1sg}.

\appendix

\subsection{Appendix A}

\noindent
The goal of this appendix is to compute the expression
$$ 
A=\frac{N_1}{2}
\int {\rm d}^3{\bf x} \; \Phi_1({\bf x})\;e^{\Phi_1({\bf x})}
+\frac{N_2}{2}
\int {\rm d}^3{\bf x}\;\Phi_2({\bf x})\;e^{\Phi_2({\bf x})}\;.
$$ 
\noindent
Using eq.(\ref{potentiel}) we obtain
\begin{eqnarray}\label{A1}
A=\frac{N_1}{2}
\int {\rm d}^3{\bf x} \; \Phi_1({\bf x})
\left[e^{\Phi_1({\bf x})}+\frac{m_2 N_2}{m_1 N_1}\;e^{\Phi_2({\bf x})}\right]
+\frac{N_2}{2} \; (a_2-\frac{m_2}{m_1}\; a _1) \;.
\end{eqnarray}
\noindent
Using eq.(\ref{Poisson}) in the spherical symmetry the first term of
eq.(\ref{A1}) becomes
\begin{eqnarray}\label{A2}
&&\frac{N_1}{2}\int {\rm d}^3{\bf x} \; \Phi_1({\bf x})
\left(e^{\Phi_1({\bf x})}+\frac{m_2 N_2}{m_1 N_1}\;e^{\Phi_2({\bf x})}\right)
=\nonumber\\
&&-\frac{N_1}{2 \eta_1} \int_0^{R_{max}} {\rm d}R  \; \Phi_1(R) \; 
\frac{{\rm d}}{{\rm d}R}(R^2 \frac{{\rm d \Phi_1}}{{\rm d}R}) \;.
\end{eqnarray}
\noindent
Using eqs. (\ref{dilatation}) and  (\ref{gauss}) we obtain
\begin{equation}\label{Phi'max}
\Phi_1^{'}(R_{max})=-\frac{\lambda}{R_{max}}(\eta_1^R +\mu \eta_2^R)\;.
\end{equation} 
\noindent
Integrating by parts (\ref {A2}) and using eqs.(\ref{dilatation}) and
(\ref{Phi'max}) we obtain
\begin{eqnarray}\label{A3}
&&\frac{N_1}{2}\int {\rm d}^3{\bf x} \; \Phi_1({\bf x})
\left[e^{\Phi_1({\bf x})}+\frac{m_2 N_2}{m_1 N_1}\;e^{\Phi_2({\bf x})}\right]
=\\
&&\frac{N_1}{2}\left(1+\frac{m_2 N_2}{m_1 N_1} \right) \; \ln{f_1}+
\frac{N_1}{2 \lambda \eta_1^R}\int_0^{\lambda} {\rm d}x \; x^2 \;
[\chi_1^{'}(x)]^2 \;.\nonumber
\end{eqnarray}
\noindent
Using eq.(\ref{Lagrange}) the second term of (\ref{A1}) yields
\begin{equation}\label{A4}
\frac{N_2}{2}\left(a_2-\frac{m_2}{m_1}a_1\right)=
\frac{N_2}{2}\left(\ln{f_2}-\frac{m_2}{m_1} \ln{f_1}\right)\;.
\end{equation} 
\noindent
Using eqs. (\ref{A3}) and (\ref{A4}) we express $A$ as
\begin{equation}\label{A5}
A=\frac{N_1}{2}\ln{f_1}+\frac{N_2}{2}\ln{f_2}+
\frac{N_1}{2 \lambda \eta_1^R}
\int_0^{\lambda} {\rm d}x \; x^2 \; [\chi_1^{'}(x)]^2\;.
\end{equation}
\noindent
We compute now 
$$ 
I(\lambda)=\int_0^{\lambda} {\rm d}x \; x^2 \; [\chi_1^{'}(x)]^2\;.
$$ 
\noindent
Using eqs.(\ref {Poissonchi}) and (\ref {champ}), we derive the function 
$$ 
B(x)=x^3 [e^{\chi_1(x)}+\mu^2 e^{\chi_2(x)}]
$$ 
\noindent
and find
\begin{equation}
\label{B'}
B^{'}(x)=-x^3 \chi_1^{'}(x) \chi_1^{''}(x)-2 x^2 [\chi_1^{'}(x)]^2
+3 \, x^2 [e^{\chi_1(x)}+\mu^2 e^{\chi_2(x)}] \; .
\end{equation}
\noindent
We integrate $B^{'}$ between $0$ and $\lambda$.
Integrating by parts the first term and using eq.(\ref{gauss}), we find
$$ 
- \frac{\lambda}{2} (\eta_1^R+\mu \;
\eta_2^R)^2+\frac{3}{2}I(\lambda)\;. 
$$ 
\noindent
The second term of eq.(\ref{B'}) yields
$$ 
-2 I(\lambda)\;.\nonumber
$$ 
\noindent
Using eq.(\ref {etai-norm}) the third term of (\ref{B'}) yields
\begin{eqnarray}
\label{term3}
3 \lambda(\eta_1^R+\mu^2\;  \eta_2^R)\;.\nonumber
\end{eqnarray}
\noindent
Hence,
$$ 
I(\lambda)=-2 \lambda^3 [e^{\chi_1(\lambda)}+\mu^2 e^{\chi_2(\lambda)}]
-\lambda (\eta_1^R+\mu \; \eta_2^R)^2+6\lambda(\eta_1^R+\mu^2 \; \eta_2^R)\;.
$$ 
\noindent
Therefore, using eqs.(\ref{etai}) and (\ref{densper}) we obtain
\begin{eqnarray}
\label{search}
\frac{N_1}{2 \lambda \eta_1^R}\int_0^{\lambda} {\rm d}x \; x^2 \;
[\chi_1^{'}(x)]^2&=&N_1 \left[ 3(1-f_1)-\frac{1}{2}
(\eta_1^R+\mu \;  \eta_2^R) \right] \\
&=&N_2 \left[ 3(1-f_2)-\frac{1}{2}
\left(\frac{1}{\mu} \; \eta_1^R+\eta_2^R\right) \right]\;. \nonumber
\end{eqnarray}
\noindent
Hence, the expression of $A$ [using eqs.(\ref{A5}) and  (\ref{search})]   is
\begin{eqnarray}\label{Afinal}
A&=&N_1 \left[\frac{1}{2}\ln{f_1}+ 3(1-f_1)-\frac{1}{2}
\left(\eta_1^R+\mu \; \eta_2^R\right) \right] \nonumber\\
&+&N_2 \left[\frac{1}{2}\ln{f_2}+ 3(1-f_2)-\frac{1}{2}
\left(\frac{1}{\mu} \; \eta_1^R+\eta_2^R\right) \right]\;. \nonumber
\end{eqnarray}

\subsection{Appendix B}

We derive here the mean field equations from the hydrostatic
equilibrium condition combined with the Poisson equation for a mixture
of two ideal gases.

The hydrostatic equilibrium condition \cite{llms}
$$
\nabla P({\vec q}) = - \left[ m_1 \, N_1 \, \rho_1({\vec q}) + m_2 \, N_2 \,
\rho_2({\vec q}) \right] \; \nabla U({\vec q})\; , 
$$
where $ P({\vec q}) $ stands for the pressure, combined with the 
equation of state for the ideal gas in local form
$$
P({\vec q}) = T \left[N_1 \, \rho_1({\vec q}) + N_2 \, \rho_2({\vec
q})\right]\; ,  
$$
yields for the particle densities
$$
 \rho_1({\vec q}) =  \rho_1^0 \; e^{ - \frac{m_1}{T} \, U({\vec q})}
 \quad , \quad \rho_2({\vec q}) =  \rho_2^0 \; e^{ - \frac{m_2}{T} \,
 U({\vec q})} 
$$
where $ \rho_1^0 $ and $ \rho_2^0 $ are constants. Inserting this
relation into the  Poisson equation 
$$
\nabla^2U({\vec q}) = 4 \pi G\, \left[ m_1 \, N_1 \, \rho_1({\vec q}) + m_2 \,
N_2 \, \rho_2({\vec q}) \right]
$$
yields in dimensionless coordinates eq.(\ref{hidro}).


\begin{thebibliography}{99}
 
\bibitem{pre} H. J. de Vega, N. S\'anchez and F. Combes, Nature, {\bf
383}, 56 (1996), Phys. Rev. {\bf D54},  6008 (1996), 
Ap. J.   {\bf 500}, 8 (1998), 	in
`Current Topics in Astrofundamental Physics: Primordial Cosmology',
NATO ASI at Erice, N. S\'anchez and A. Zichichi editors, vol 511, Kluwer, 1998.

\bibitem{pcm} D. Pfenniger, F. Combes, L. Martinet, A\&A {\bf 285}, 79 (1994),
D. Pfenniger, F. Combes,  A\&A {\bf 285}, 94 (1994)  

\bibitem{hidro} R. Emden, Gaskugeln, Teubner, Leipzig und Berlin, 1907. 

S. Chandrasekhar, `An introduction to the Study
of Stellar  Structure', Chicago Univ. Press, 1939. 

V. A. Antonov, Vest. Leningrad Univ. 7, 135 (1962).

D. Lynden-Bell and R Wood, Mon. Not. R. astr. Soc. {\bf 138}, 495 (1968).

T. Padmanabhan, Phys. Rep. 188, 285 (1990).

G. Horwitz and J. Katz, Ap. J. {\bf 211}, 226 (1977) and 
{\bf 222}, 941 (1978). 

W. C. Saslaw, 
`Gravitational Physics of stellar and galactic systems',
Cambridge Univ. Press, 1987.

J. Binney and S. Tremaine, Galactic Dynamics, Princeton
Univ. Press.

\bibitem{1sg} H.J. de Vega, N. S\'anchez, Phys. Lett. {\bf B490}, 180
(2000), Nuclear Physics {\bf B 625}, 409 and  460 (2002).

\bibitem{Lipatov} L. N. Lipatov, JETP {\bf 45}, 216 (1977).

\bibitem{llms} L. D. Landau and E. M. Lifchitz, Physique Statistique,
4\`eme \'edition, Mir-Ellipses, 1996.

L. D. Landau and E. M. Lifchitz, M\'ecanique des Fluides, \'Editions
Mir, Moscou 1971. 
\end{thebibliography}
\end{document}